%% file: paper4.tex
\documentclass[prb,aps,floatfix, reprint, twocolumn, superscriptaddress, titlepage,amssymb, amsmath]{revtex4-1}

\usepackage{graphicx, times, epsfig, color, mathtools}
\def\be{\begin{equation}}
\def\ee{\end{equation}}
\def\bea{\begin{eqnarray}}
\def\eea{\end{eqnarray}}
\def\etal{{\it et al.}}
\def\deg{$^{\circ}$}
\def\asih{{\it a}-Si:H}
\def\asi{{\it a}-Si}

\def\kb{\mathbf k}

\begin{document} 
\title{
Small-angle X-ray scattering in amorphous silicon: A computational study}

\author{Durga Paudel}
\email{durga.paudel@usm.edu} 
\affiliation{Department of Physics and Astronomy, 
The University of Southern Mississippi, Hattiesburg, MS 39406}

\author{Raymond Atta-Fynn} 
\email{r.attafynn@uta.edu} 
\affiliation{Department of Physics, University of Texas, Arlington, TX 76019}

\author{David A. Drabold} 
\email{drabold@ohio.edu}
\affiliation{Department of Physics, Ohio University, Athens, Ohio 45701}

\author{Stephen R. Elliott} 
\email{sre1@cam.ac.uk}
\affiliation{Department of Chemistry, University of Cambridge, CB2 1EW, 
Cambridge, United Kingdom}

\author{Parthapratim Biswas}
\email[Corresponding author:\,]{partha.biswas@usm.edu}
\affiliation{Department of Physics and Astronomy, 
The University of Southern Mississippi, Hattiesburg, MS 39406}

\begin{abstract}
We present a computational study of small-angle X-ray scattering (SAXS) 
in amorphous silicon ({\asi}) with particular emphasis on the morphology and 
microstructure of voids.  The relationship between the scattering 
intensity in SAXS and the three-dimensional structure of 
nanoscale inhomogeneities or voids is addressed by generating ultra-large 
high-quality {\asi} networks with 0.1-0.3 \% 
volume concentration of voids, as observed in experiments using 
SAXS and positron annihilation spectroscopy.  A systematic study 
of the variation of the scattering intensity in the small-angle scattering 
region with the size, shape, number density, and the spatial 
distribution of the voids in the networks is presented. 
Our results suggest that the scattering intensity in the 
small-angle region is particularly sensitive to the size 
and the total volume-fraction of the voids, but the effect 
of the geometry or shape of the voids is less pronounced 
in the intensity profiles. A comparison of the average size 
of the voids obtained from the simulated values of the 
intensity, using the Guinier approximation and Kratky plots, 
with those from the spatial distribution of the 
atoms in the vicinity of void surfaces is presented.  
\end{abstract}
\maketitle

\section{Introduction}

Small-angle X-ray scattering (SAXS) is a powerful method for 
studying structural inhomogeneities on the extended length 
scale in solids and condensed-phase systems in 
solution.~\cite{Guinier-book, Svergun-book, Hura2009}  
While X-ray crystallography and 
nuclear magnetic resonance (NMR) spectroscopy can provide high-resolution 
structural information, small-angle scattering of X-rays and neutrons 
is particularly useful in probing low-resolution structural 
characteristics of partially-ordered and disordered objects on 
the nanometer length scale, which is often complemented with 
results from X-ray diffraction and NMR measurements.~\cite{Mertens2010}  
Since its first inception by Guinier~\cite{Guinier-book} 
in the late 1930s, SAXS has been employed extensively in probing 
structural properties of a variety of crystalline and 
non-crystalline solids, including nanocomposites, alloys, glasses, ceramics, and 
polymers.~\cite{Guinier-book, Svergun-book, Glatter1977b} In recent 
years, the advancement of SAXS instrumentation and the availability 
of high-brilliance X-ray sources have led to the development and 
emergence of SAXS as a principal tool in structural biology~\cite{Grant2011, Perez2012} 
for studying an array of biological objects ranging from large 
macromolecules~\cite{Putnam2007}, biopolymers,~\cite{Hyland_2013} RNA 
folding,~\cite{Pollack_2011, Doniach2001} multi-domain proteins with flexible linkers,~\cite{Bernado2007} and 
intrinsically disordered proteins.~\cite{Bernado2011}
In spite of the tremendous success and the widespread applications of SAXS in 
obtaining structural information on the size, shape, and 
compactness of the scattering objects (e.g., macromolecules 
in solution or voids in amorphous environments), a direct 
determination of the three-dimensional structure of the 
scatterers solely based on the information content of a given SAXS 
data set is impossible unless additional independent information is
available to complement the SAXS data.  Since the distribution of the 
scatterers produces a rotational averaging of the intensity 
in reciprocal space, the absence of directional (or phase) 
information between the scatterers makes it extremely 
difficult to unambiguously reconstruct the three-dimensional shape of a mono-disperse 
scattering object from one-dimensional intensity profiles. While 
the problem is more acute for poly-disperse objects in biomolecular 
systems, the analysis of SAXS data in structural biology is 
often accompanied by complementary structural information 
from high-resolution X-ray crystallography and NMR 
data, providing additional information on the structure 
of the constituents or sub-units of the scattering objects 
in order to develop a three-dimensional model.~\cite{Zheng2002} 
Complications also arise in interpreting and translating 
experimental SAXS data from the reciprocal-space domain 
to the real-space domain owing to the finite size of the 
data set, sampled only at specific points in reciprocal 
space.  In an authoritative treatment, Moore~\cite{Moore1980} 
has addressed this problem by developing a framework based 
on the sampling theorem of Shannon,~\cite{Shannon-book} which 
provides an elegant ansatz to extract the full information 
content in a given data set and to estimate the errors 
associated with the parameters derived from the analysis. 

Given the complexity involved in the analysis of experimental 
SAXS data and the subsequent determination of a three-dimensional 
model of the scattering objects, a natural approach to address 
the problem is to study the relationship between the SAXS 
intensity and the structure of scattering objects by directly 
simulating the scattering intensity from realistic model 
configurations, obtained from independent calculations.  
In this paper, we address the morphology of voids in {\asi} with particular emphasis
on the relationship between the (simulated) intensity from SAXS and the 
shape, size, density, and the spatial distribution of the 
voids in amorphous silicon.  
While the problem has been studied extensively using experimental 
SAXS data for {\asi} and {\asih},\cite{Will-JNC-1989, Mahan1991, Mahan1989, Acco1996} there exist 
only a few computational studies~\cite{Biswas-1989, Ben2011} 
that have attempted to address the problem from an atomistic 
point of view using rather small models of {\asi}, 
containing only 500 to 4000 atoms. Since the information that 
resides in the small-angle region of reciprocal space is connected to 
real space via the Fourier transformation, it is necessary to have a significantly large model 
to include any structural correlations that may 
originate from distant atoms in order to produce 
the correct long-wavelength behavior of the scattering intensity. 
Thus, accurate simulations of SAXS in non-crystalline solids 
were hampered in the past by the lack of appropriately large 
structural models of {\asi}, with a linear size of several tens of angstroms, 
which are necessary for reliable computation of the scattering 
intensity in the small-angle region. 

We should mention that an impressive number of computational 
and semi-analytical studies can be found in the literature from the 
past decades that address the relationship between the scattering 
intensity in SAXS and the morphological characteristics of 
inhomogeneities present in a sample, using the 
homogeneous-medium approximation.~\cite{Guinier-book, Letcher_1966, Vrij_1989, Jemian_2009} 
Such an approach, however, 
crucially relies on the assumption that the length scale 
($l$) associated with the inhomogeneities is 
significantly larger than the atomic-scale structure ($R$) 
of the embedding medium (i.e., $l >> R$), so that any 
density fluctuations that may originate from the atomic-scale structure 
of the embedding matrix on the length scale of $R$ can 
be neglected for the computation of the intensity in the 
relevant small-angle region of interest. It thus 
readily follows that, given the length scale of the voids 
in {\asi} ($l \approx$ 10--18 {\AA}) and the atomic-scale 
structure of the amorphous-silicon matrix ($R\approx$ 10--15 {\AA}), 
neither the homogeneous-medium approximation 
nor an approach based upon relatively small atomistic models 
of {\asi}, consisting of 500--4000 Si atoms, 
is adequate for accurate simulations of SAXS intensity in 
the presence of nanometer-size inhomogeneities in 
amorphous silicon. 

The importance of atomistic simulations becomes particularly 
apparent in determining the effect of surface relaxation 
on the shape of the inhomogeneities and its possible 
manifestation on SAXS intensities, which cannot be 
addressed realistically using the homogeneous-medium 
approximation. Furthermore, the behavior of the static 
structure factor in the small-angle limit is by itself 
an important topic for studying the long-wavelength 
density fluctuations in disordered systems. 
In an influential paper 
appearing in the Proceedings of the National Academy 
of Sciences, Xie {\etal}~\cite{Xie2013} presented highly sensitive
transmission X-ray scattering data of {\it a}-Si
samples to examine the infinite-wavelength limit ($q \to 0$) of
the structure factor $S(q)$ for determining the degree of
hyperuniformity, and reported a value of $S(0) = 0.0075 \pm 0.0005$.
Following these authors, $S(q \to 0)$ can be used as a
figure-of-merit to study the quality of the amorphous-silicon
network generated in our simulations. Here, we shall show that 
the value of $S(q\to 0)$ obtained from our simulations is 
closer to the experimental value than the computed value 
reported in the literature by de Graff and 
Thorpe.~\cite{Graff-2010} For a discussion on 
hyperuniformity and its applications to disordered 
systems, the readers may refer to the work by 
Torquato and co-workers.~\cite{Tor1, Tor2}

The remainder of the paper is as follows. In Sec.\,II, we address the
computational method associated with the production of ultra-large 
high-quality structural models of {\asi}, which is followed by 
the calculation of the SAXS intensity and the construction of voids of different 
shapes, sizes, densities, and their spatial distributions in 
several model configurations of amorphous silicon.  Section III 
discusses the results from our simulations where we address the 
characteristic structural properties of the models and compare 
the simulated structure factor with the high-resolution 
structure-factor data of {\asi} from experiments. 
This is followed by a discussion on the restructuring of a 
void surface upon total-energy relaxation and the subsequent 
changes in the shape and topology of the surface atoms. 
Thereafter, we examine the relationship between the 
morphology of the voids and the scattering intensity in SAXS, 
by studying several models of {\asi} with a varying size, shape, and concentration 
of the voids.  
A comparison of the size of the voids with the same obtained 
from the simulated intensity in the small-angle region is 
also presented from Guinier and Kratky plots.  
Section IV presents the conclusions of our work. 

\section{Computational Methods}
\subsection{Large-scale modeling of {\asi} for simulation of SAXS}

Since the main purpose of the present work is to study the structure 
and statistical properties of extended-scale inhomogeneities on the 
nanometer length scale, we are interested in the scattering region 
associated with small wave vectors in the range of 0--1 {\AA}$^{-1}$. 
For inhomogeneities, such as voids, with a typical size of 
$l \approx$ 10--20 {\AA}, one needs to measure scattering intensities 
for the wave vectors in the vicinity of $k =  2\pi/l \approx$ 0.3--0.6 {\AA}$^{-1}$. This 
means that the appropriate structural models needed to be used 
in the simulation of small-angle X-ray scattering must have a 
linear dimension of several nanometers in order to compute 
statistically-reproducible physical quantities from the simulated 
SAXS data. To fulfill this requirement, we generated ultra-large atomistic 
configurations of {\asi} using classical molecular-dynamics (MD) 
simulations, as described below. 

Two independent initial configurations, each comprising $N=262,400$
Si atoms, were generated by randomly placing atoms in a cubic simulation  
box of length 176.12 {\AA}, so that the minimum distance between 
any two Si atoms was 2.0 {\AA}. This corresponds to 
a mass density of 2.24 g/cm$^3$ 
for the models, which is identical to the experimental mass 
density of {\asi} reported by Custer {\etal}~\cite{Custer1994}  
Starting from these initial configurations, MD simulations were 
carried out in the canonical ensemble by describing the interatomic 
interaction between Si atoms using the 
modified Stillinger-Weber potential.\cite{sw1,sw2} 
The equations of motion were integrated using the velocity-Verlet 
algorithm with a time step of $\Delta t=1$ fs  and the Nos{\'e}-Hoover 
thermostat\cite{nose,hoover,chains} was employed to control the 
simulation temperature, with a thermostat period of $\tau=0.2$ ps.  
The initial temperature of each configuration was set to 1800 K and 
the configurations were equilibrated for 20 ps.
After equilibration at 1800 K, each configuration was cooled 
to 300 K over a total time period of 300 ps with a cooling 
rate of 5 K/ps.  Since atomistic models of amorphous 
silicon obtained from MD simulations, using a single heating-and-cooling 
cycle, cannot produce good structural properties owing to the large 
volume and dimensionality of the phase space in a limited simulation time, 
we repeated the heating-and-cooling cycles 30 times in order 
to sample the phase space extensively for producing high-quality atomistic 
configurations with excellent structural properties. For the present 
simulations, this translates into a total simulation 
time of 9 nanoseconds for each configuration.  The final configurations 
were obtained by minimizing the 
total energy with respect to the atomic positions 
using the limited-memory BFGS algorithm.\cite{bfgs1,bfgs2} 
In the following, we refer to these final configurations 
as M-1 and M-2, and we have used them for further simulation 
and analyses of the scattering intensity in SAXS. The 
characteristic structural properties of these models are 
listed in Table \ref{TAB2}.

\subsection{Simulation of SAXS intensity for amorphous solids}

\begin{figure}[t!]
\includegraphics[height=2.2 in, width=0.4\textwidth]{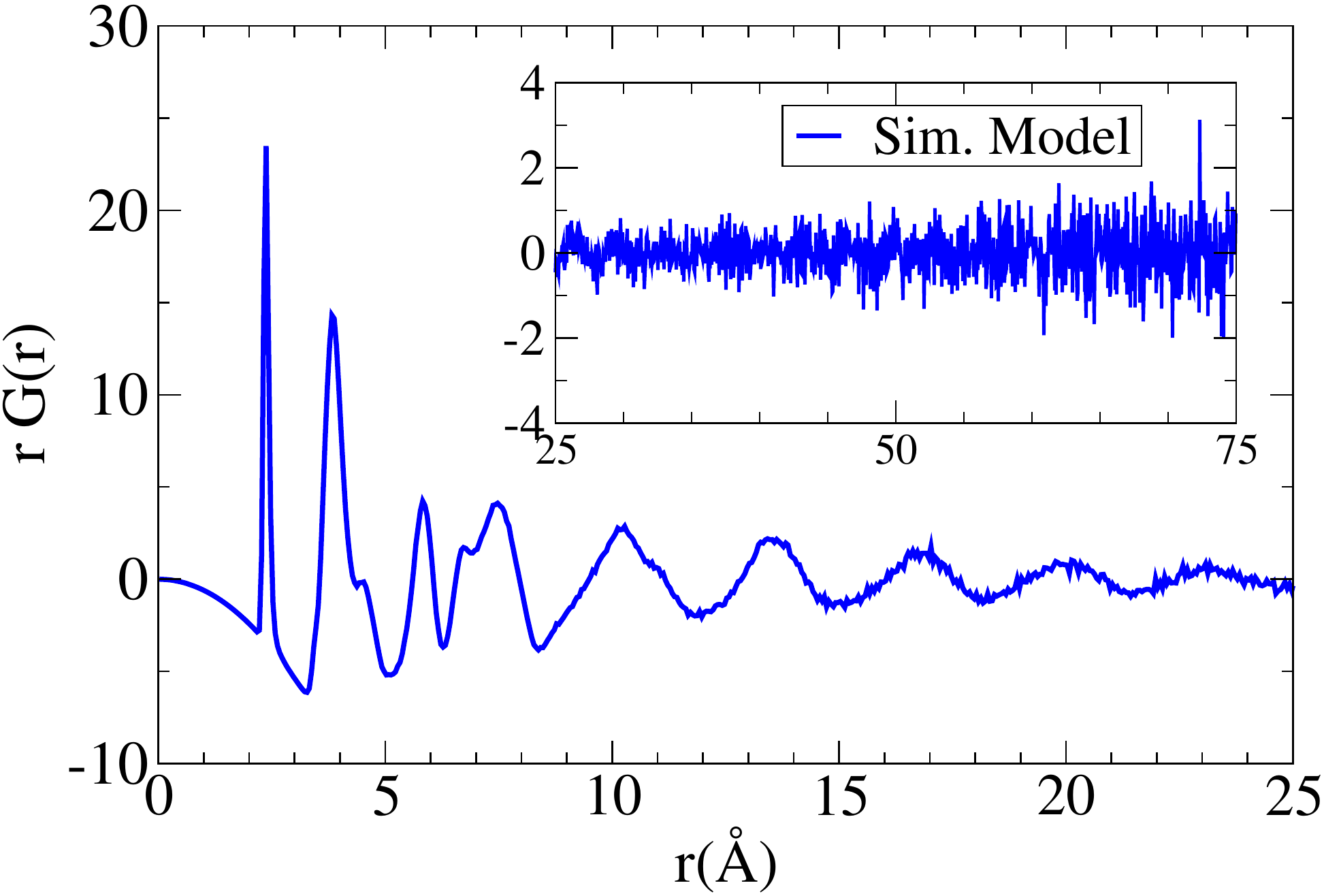}
\caption{
The variation of $r\,G(r)$ with $r$ for the M1 model of {\asi} 
containing 262,400 atoms. The inset shows the growing 
fluctuations in $r\,G(r)$ beyond 40 {\AA}, which affect 
the evaluation of the integral in Eq.\,\ref{gr2}. See Sec. 
IIB for a discussion. 
}
\label{FIG1}
\end{figure}

For disordered and amorphous systems, the intensity of X-ray scattering 
is a function of the microscopic state of the system. 
The scattering intensity depends on the individual scattering units 
(e.g., atoms, molecules, cells) and the characteristic 
statistical distribution of the units in the system.  
The scattering intensity for a system consisting of $N$ 
atoms can be written as, 
\be 
I({\kb}) = \sum_i^N \sum_j^N f_i({\kb}) f_j({\kb}) \exp[\imath {\kb} \cdot (\mathbf r_i - \mathbf r_j)], 
\label{xray}
\ee 
\noindent 
where the contribution from an individual atom enters through 
the atomic form-factor $f_i({\kb})$ and the structural information 
follows from the (positional) distribution of the constitutent 
atoms in the system. Here, the wave-vector transfer, ${\kb}$, 
is the difference between the scattered 
($\mathbf k_f$) and incident ($\mathbf k_i$) wave vectors, and 
its magnitude is given by $k = |\mathbf k_f-\mathbf k_i| 
= 4\pi \sin\theta/\lambda$, where $2\theta$ and $\lambda$ 
are the scattering angle and the wavelength of the 
incident X-ray radiation (e.g., 1.54 {\AA} for the 
Cu K$_{\alpha}$ line), respectively.  
While Eq.\,(\ref{xray}) 
can be evaluated directly for small systems, it is 
computationally very demanding and infeasible to compute 
the intensity for large models with hundreds of 
thousands of atoms.  Since it is necessary to minimize 
surface effects by imposing the periodic boundary 
conditions, one needs to evaluate the double 
sum in Eq.\,(\ref{xray}) in order to compute the 
intensity values. 
Further, the computation of the configurational-averaged 
values of the scattering intensity, for a given $k$, requires angular 
averaging over all possible directions of ${\kb}$ over a 
solid angle of $4\pi$.  
Finally, using the well-known sampling theorem of 
Shannon,~\cite{Shannon-book} it can be shown that,  
in order to extract the full information content 
of SAXS data, one must sample the scattering intensity 
at equally-spaced points, $k_i$, with spacing 
$\Delta k$ -- also known as Shannon channels -- 
such that $\Delta k \le \pi/l$, where $l$ is the 
maximum linear size of the inhomogeneities 
dispersed in the system.~\cite{Moore1980, Damaschun1968}  
These considerations lead to the conclusion that, for 
a system with $10^5$ atoms, one requires to compute 
approximately 10$^{15}$ or more operations in order to obtain 
the intensity plot from Eq.\,(\ref{xray}). 
The conventional approach is to carry out the averaging 
procedure analytically by introducing a 
pair-correlation function $g(r)$, which is associated with 
the probability of finding an atom at a distance $r$, given 
that there is an atom at $r = 0 $. By invoking the assumptions 
that the system is homogeneous and isotropic and that the 
strong peak near $k$=0, originating from a constant density 
term, does not provide any structural information and thus 
can be removed from consideration, one arrives at the 
following expression for the scattering intensity for a monatomic 
system, 
\be 
I_N(k) = N f^2(k)\,S(k), 
\label{gr1}
\ee 
where
\bea 
S(k) & = & 1 + \frac{4\pi\rho}{k} \int_0^{\infty} r (g(r) - 1) \,\sin{kr}\,dr \nonumber \\
& \approx & 1 + \int_0^{R} r \,G(r)\,\frac{\sin{kr}}{kr}\,dr. 
\label{gr2}
\eea 
In Eq.\,(\ref{gr2}), we have introduced the reduced distribution function, 
$G(r) = 4 \pi \rho\, r(g(r) - 1)$. For computational purposes, it is also 
necessary to replace the upper limit of the integral by a large but finite 
cutoff distance, $R$, beyond which $(g(r)-1)$ tends to vanish. For 
finite-size models, the cutoff distance, $R$, is generally, but not 
necessarily,  chosen to be 
the half of the box length for a cubic model of linear size $L$. Equation (\ref{gr2}) 
can be readily employed to compute the structure factor reliably in 
the wide-angle limit but the difficulty remains for very small 
values of $k$. It has been shown by Levashov {\etal}\cite{Levashov2005} 
that $g(r)$ converges to unity very slowly, and at finite temperature 
there exist small but intrinsic fluctuations, even for a very 
large value of $R$. 
In the small-angle limit, the term $\sin(kr)/kr$ in Eq.\,(\ref{gr2}) 
changes very slowly but the fluctuations in $r\,G(r)$ grow 
considerably beyond a certain radial distance $R_c$ due to 
the presence of the $r^2$ term. 
Thus, $R_c$ must be as large 
as possible to extract structural information for small 
$k$ values.  It is often convenient to write Eq.\,(\ref{gr2}) 
in two parts by introducing a damping factor $\gamma(r)$ in the 
region $r \ge R_c$.  The resulting equation now reads, 
\be 
S(k) \approx 1+\int_0^{R_c} rG(r)\frac{\sin{kr}}{kr} dr 
+ \int_{R_c}^{R} \gamma(r) rG(r)\frac{\sin{kr}}{kr} dr. 
\label{gr3}
\ee
\noindent 
Computational studies on $G(r)$ in {\asi}, using large 
simulated models, indicate that the optimum value of 
$R_c$ is of the order of 30--40 {\AA}. Beyond this distance, it 
is difficult to distinguish $G(r)$ from numerical noise and the 
accuracy of the integral in Eq.\,(\ref{gr2}) is found to be 
affected by the presence of growing oscillations 
in $rG(r)$.  To mitigate the effect of the truncation of the 
upper limit of the integral at small $k$ 
values, we have used an exponential damping factor, 
$\gamma(r)=\exp[-(r-R_c)/\sigma]$, in the region 
$r \ge R_c$.  Numerical experiments indicate that a choice of 
$R_c$ = 35--40 {\AA} and $\sigma$ = 1 {\AA} is appropriate 
for our models. Since structural information on 
extended-scale inhomogeneities generally resides beyond the first 
few neighboring shells, this observation implies that,  
even with very large models, one must be careful to interpret the 
simulated values of the scattering intensity below 
$k = 2\pi/R_c \approx 0.1$ {\AA}$^{-1}$ due to a low signal-to-noise 
ratio in $rG(r)$, as shown in Fig.\,\ref{FIG1}. 
Once the structure factor is available, the 
reduced scattering intensity, $I(k)$, can be obtained from 
the expression, 
\be
I(k)=\frac{I_N(k)}{N} = f^2(k) S(k),
\label{EQ6}
\ee
\noindent 
where $N$ is the number of atoms in the model. The atomic 
form-factor can be obtained from the International Tables 
for Crystallography\cite{Int-table1993} or from a suitable 
approximated form of $f(k)$.\cite{Doyle1968,Smith1962}
At a finite temperature $T$, the expression for the reduced intensity 
in Eq.\,(\ref{EQ6}) is multiplied by the Debye-Waller (DW) 
factor,~\cite{Debye1913, Waller1923} $\exp(-2M)$, where 
$M=(8\pi^2\sin^2{\theta}/\lambda^2)(u^2/3)$ and $u^2(T)$ is the mean-square 
displacement of Si atoms in the amorphous state at temperature $T$. The 
Debye-Waller-corrected reduced intensity can be written as, 
\be 
I_{DW}(k, T) = \exp(-2M)\, I(k).  
\label{EQ77}
\ee 
The calculation of the Debye-Waller factor for the amorphous state 
is, by itself, an interesting problem and it is related to the 
vibrational dynamics of the atoms at a given temperature.  The 
factor plays an important role in extracting structural information 
from X-ray scattering data by reducing and redistributing the 
scattering intensity at high temperature. At room temperature, 
the DW factor affects the intensity values only marginally for small 
values of $k$ and it can be replaced by unity for the computation of 
scattering intensity in the region $k < $ 1.0 {\AA}$^{-1}$.  

\subsection{Geometry of voids in {\asi} for SAXS simulation} 
\begin{figure}[t!]
\includegraphics[height=1.5 in, width=0.23\textwidth]{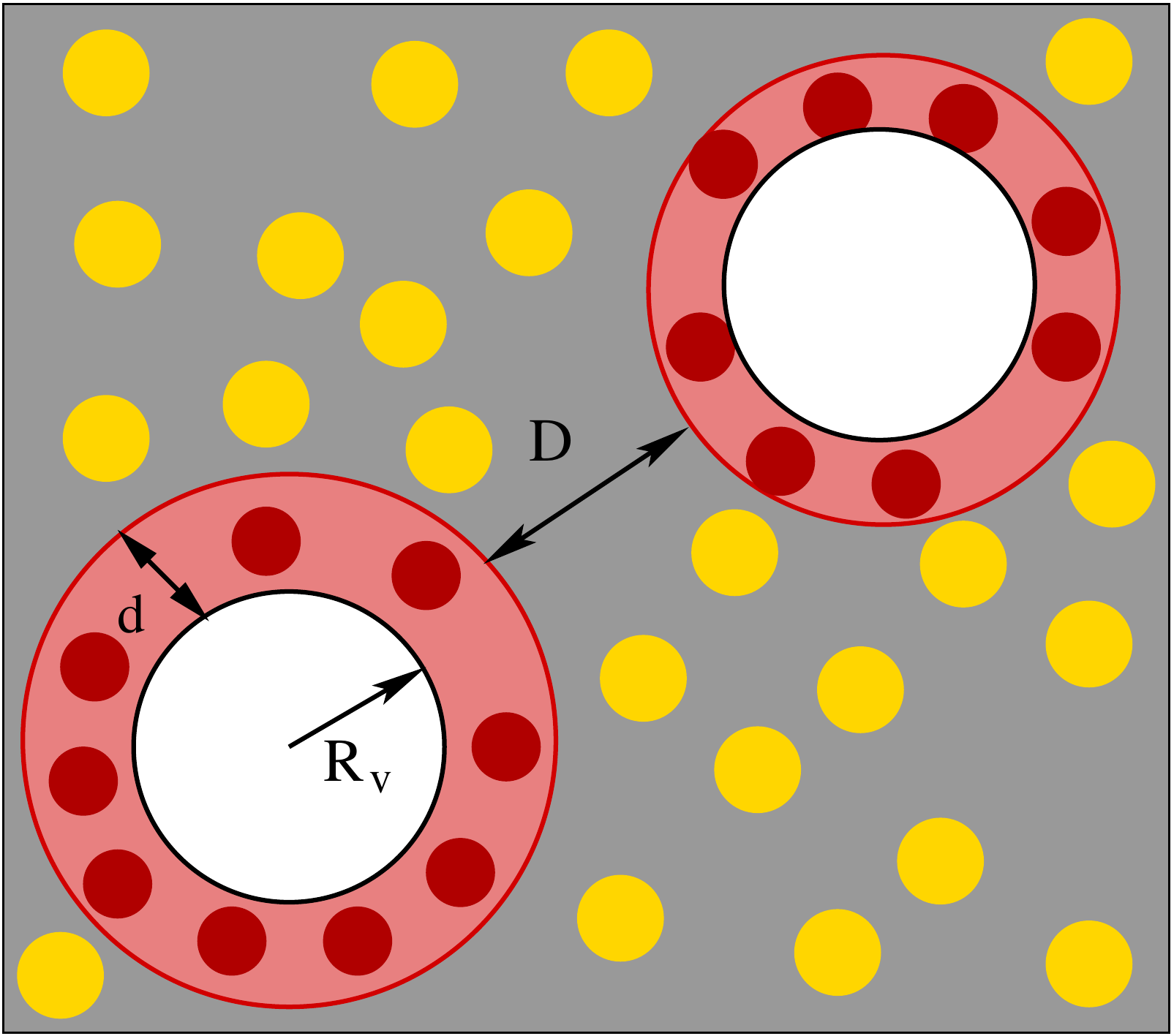}
\includegraphics[height=1.5 in, width=0.23\textwidth]{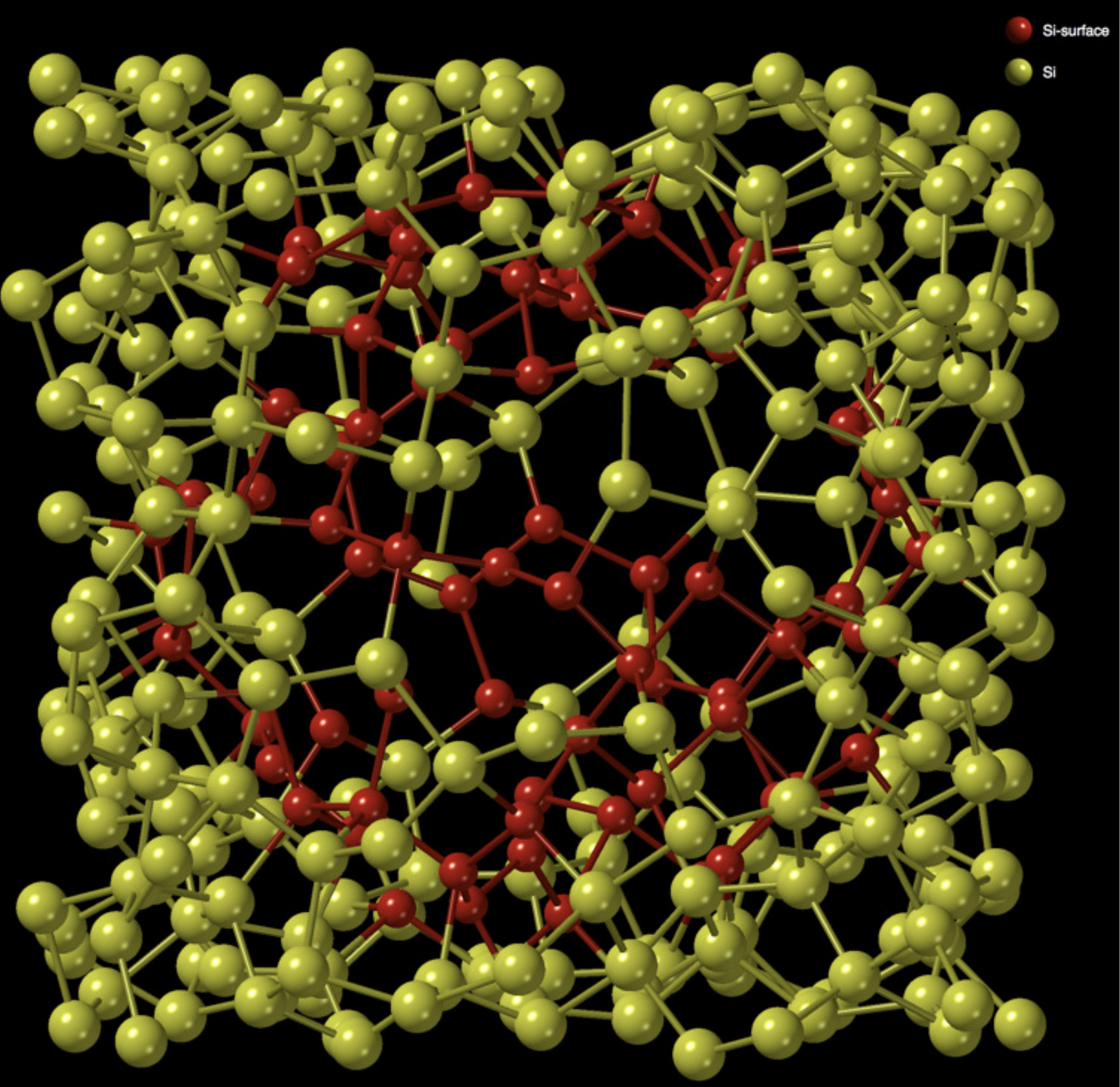}
\caption{
A schematic representation of voids in two dimensions (left)
showing the characteristic lengths associated with void size ($R_v$),
interface width ($d$), and the surface-to-surface distance
($D$) between two voids.  The figure on the right-hand side
shows a spherical void of radius 6 {\AA} in a network of
size 10 {\AA}. For visual clarity, the silicon atoms on
the void surface, having an interface width of $d$ = 2.8 {\AA},
and the bulk region are shown in red and yellow colors,
respectively.
}
\label{FIG2}
\end{figure}

In order to examine the relationship between the morphology of 
voids and the intensity of the small-angle X-ray scattering in {\asi}, 
it is necessary to construct a variety of void distributions 
in {\asi} networks,  which are characterized by different shapes, sizes, 
and number densities of voids. Since experimental data from 
IR, NMR, SAXS,~\cite{NREL411, Mahan1989, Chabal1987, Will-JNC-1989, Mahan1989_solar}
positron annihilation spectroscopy (PAS),~\cite{Muramatsu1994, Pas2016, Smets2017} 
and implanted helium-effusion measurements~\cite{Beyer2004, Beyer2011} 
suggest that the percentage of 
void-volume fraction ($f_v$) in {\asi} and {\asih} varies from 
0.1\% to 0.3\% of the total volume of the samples, and the 
typical size or radius of the voids ranges from 
5 {\AA} to 10 {\AA}, we have restricted ourselves to generating
structural models of {\asi} with voids that simultaneously 
satisfy both the requirement of void-volume fraction 
and the size of the voids.  
Toward that end, we have created several void distributions, which are 
characterized by spherical, ellipsoidal, and cylindrical voids, 
by randomly generating 
void centers within two model networks, M-1 and M-2, consisting 
of 262,400 Si atoms in a cubic simulation cell of length 
176.12 {\AA}. To ensure that the randomly-generated void 
distributions in the networks are as realistic as one observes 
in experiments, we introduced three characteristic lengths, 
$R_v$, $d$, and $D$, as illustrated in Fig.\,\ref{FIG2}. 
The radius of a spherical void is given by $R_v$, whereas $d$ 
indicates the width of the spherical concentric region between radii 
$R_v$ and $R_v+d$, which determines the interface region of the (spherical)
void and the bulk network. Silicon atoms in this 
region will be referred to as interface atoms, and we shall see 
later that these atoms play an important role in the relaxation 
of void surfaces. The atoms within a void region are removed 
from the system in order to produce an empty cavity or a void. 
$D$ indicates the minimum interface-to-interface 
distance between two neighboring voids, as shown in Fig.\,\ref{FIG2}. 
This implies that the 
center-to-center distance, $r_{ij}$, between two spherical 
voids at sites $i$ and $j$ satisfies the constraint 
$r_{ij} \ge 2(R_v+d) + D$. By choosing appropriate values of 
$f_v$, $R_v$, and $D$, one can produce a variety of 
void distributions, which are consistent with experimental 
results as far as the void-volume fraction and the size 
of the voids are concerned. 
For example, by choosing a large (or small) 
value of $D$, one can construct a sparse (or clustered) 
distribution of voids. Throughout the study, we have used $d$ = 2.8 {\AA} that 
corresponds to the maximum nearest-neighbor distance 
between two silicon atoms in {\asi}. For a given set of 
$f_v$, $R_v$, $D$, and the shape of the voids, one can 
compute the number of voids $n_v = f_vV/\nu$, where $\nu$ 
and $V$ are the volumes associated with an individual 
void and the simulation cell, respectively. 
For non-spherical voids, such as ellipsoidal and cylindrical 
voids, we replace $R_v$ by appropriate lengths $R_v^e$ and 
$R_v^{cy}$, which indicate the geometric mean radius of an 
ellipsoidal void and the cross-sectional radius of a 
cylindrical void, respectively.  Ellipsoidal voids were generated 
by constructing triaxial ellipsoids with the axes ratios 
$a:b:c$ = $\frac{R_v}{2}:R_v:2R_v$, so that the geometric 
mean radius $R_v^e$ (=$\sqrt[3]{abc}$) is equal to 
the radius $R_v$ of a spherical void for a given $f_v$.
For cylindrical voids, the height of a cylinder was taken 
to be three times its cross-sectional radius, $R_{cy}$, and 
the latter was chosen so that the volume of the cylinder was 
identical to that of a sphere or an ellipsoid (see Ref.\,\onlinecite{Krat-note}). 
The orientations of the ellipsoidal and cylindrical voids 
were randomly generated by constructing a three-dimensional 
unit random vector from the center of each void and aligning 
the major axis of an ellipsoid or a cylinder along that 
direction. An example of a spherical void of radius $R_v$ = 6 {\AA} 
and interface width of $d$ = 2.8 {\AA} is shown in Fig.\,\ref{FIG2}, 
which is embedded in a region of the {\asi} network 
of linear dimension 10 {\AA}. The silicon atoms in the bulk 
and interface regions of the void are shown in yellow 
and red colors, respectively. 

\begin{table}[t!] 
\caption{\label{TAB3}
Characteristic properties of {\asi} models with void 
distributions used in this work. $R$, $N_b$, $N_s$,
and $N_v$ indicate the actual radius, and the total 
number of bulk, surface, and void atoms, respectively. The 
percentage void-volume fraction ($f_v$), number 
density per cm$^3$ ($n_\rho$), and the average radius 
of gyration ($R_g$) of the voids are listed. See 
text for the nomenclature of the models listed 
below. 
}
\begin{ruledtabular}
\begin{tabular}{lcrrccccc}
Model &$R$ (\AA)   & $N_b$   & $N_{v}$     & $N_{s}$  & $f_{v}$ &  $n_{\rho} \times10^{19}$
& $R_g$ (\AA) \\
\hline
SP6-R6    &6.0       & 261584   & 259   & 557         & 0.1   & 0.11           & 6.13 \\
SP3-R8    &8.0       & 261634   & 306   & 460           & 0.1   & 0.05           & 8.09 \\
SP12-R6   &6.0       & 260761   & 533   & 1106                & 0.2   & 0.22           & 6.13 \\
SP5-R8    &8.0       & 261126   & 508   & 766         & 0.2   & 0.09           & 8.09 \\
SP18-R6   &6.0       & 259936   & 801   & 1663                & 0.3   & 0.33           & 6.13 \\
SP8-R8    &8.0       & 260371   & 819   & 1210          & 0.3   & 0.15         & 8.09 \\
\hline
EL6-R6    &6.0       & 261491   & 260   & 649&        0.1     & 0.11   & 7.3 \\
EL3-R8    &8.0       & 261563   & 302   & 535&        0.1     & 0.05   & 9.66 \\
EL12-R6   &6.0       & 260578   & 502   & 1320&      0.2     & 0.22  & 7.31 \\
EL5-R8    &8.0       & 261005   & 513   & 882&        0.2     & 0.09   & 9.66 \\
EL18-R6   &6.0       & 259666   & 763   & 1971&      0.3     & 0.33   & 6.15 \\
EL8-R8    &8.0       & 260173   & 825   & 1402&      0.3     & 0.15  & 9.66 \\
\hline
CY6-R5&  4.58        & 261731   & 260   &409          & 0.1   & 0.11   & 5.83 \\
CY3-R6&  6.10        & 261752   & 298   & 350         & 0.1   & 0.05   & 7.73 \\
CY12-R5& 4.58        & 261065   & 511   & 824         & 0.2   & 0.22   & 5.78 \\
CY5-R6 & 6.10        & 261327   & 494   & 579         & 0.2   & 0.09   & 7.74 \\
CY18-R5& 4.58        & 260389   & 774   & 1237       & 0.3   & 0.33   & 5.8 \\
CY8-R6&  6.10        & 260696   & 792   & 912         & 0.3   & 0.15   & 7.75 \\
\hline
SP18-D1-R6&6.0        & 259935   & 789   & 1676            & 0.3     & 0.33  & 6.13 \\
SP18-D8-R6&6.0        & 259948   & 785   & 1667            & 0.3     & 0.33  & 6.13 \\
SP18-D14-R6&6.0       & 259949   & 795   & 1656            & 0.3     & 0.33  & 6.09 \\
\end{tabular}
\end{ruledtabular}
\end{table}

Table \ref{TAB3} lists some characteristic features of
voids and the resulting models obtained by incorporating 
voids of different shapes, sizes, numbers, and void-volume 
fractions.  In order to produce a statistically-significant
number of voids for a given volume fraction of voids, the 
radii of the voids were restricted to 5--8 {\AA}. For $f_v$ 
= 0.1\%, 0.2\%, and 0.3\%, spherical, ellipsoidal and 
cylindrical voids of different sizes were generated 
randomly within the networks in such a way that none 
of the voids was too close to the boundary of the networks. 
In this work, we have studied a total of 21 models that are 
listed in column 1 of Table \ref{TAB3}.  Each of the models 
is indicated by its shape, the number of voids present in 
the model, and the approximate linear size of the voids.  
For example, EL6-R6 indicates a model with 6 
ellipsoidal voids of radius 6 {\AA}. Similarly, SP18-D8-R6 
implies a model with 18 spherical voids of radius 
6 {\AA}, which are separated by the surface-to-surface distance 
($D$) of at least 8 {\AA}.  For cylindrical voids, the exact 
value of the cross-sectional radius of a void is given 
in column 1 of Table \ref{TAB3}.  The total number of bulk 
($N_b$), surface ($N_s$), and void~\cite{void_atom} ($N_v$) 
atoms, along with the corresponding void-volume fraction 
($f_v$), number density of voids per cm$^{3}$ ($n_\rho$), 
and the average radius of gyration ($R_g$) of the voids 
for each model after total-energy relaxation are also listed 
in Table \ref{TAB3}. The average radius of gyration, $R_g$, 
of voids in a model configuration can be obtained from the 
atomic coordinates of all the interface atoms in a model. 

\section{Results and Discussion}

In the preceding sections, we have seen that the structural 
information from extended length scales chiefly resides 
in the small-angle scattering region of wave vectors, 
$k \le$ 1.0 {\AA}$^{-1}$.  In view of our earlier observation 
that the computed values of the structure factor could be 
affected by finite-size effects, owing to the growing 
oscillations in $rG(r)$ at large $r$, it is 
necessary to examine the accuracy of the simulated values 
of the scattering intensity before addressing the 
relationship between the scattering intensity and the 
inhomogeneities or voids from SAXS measurements.
\begin{figure}[t!] 
\centering
\includegraphics[width=0.4\textwidth]{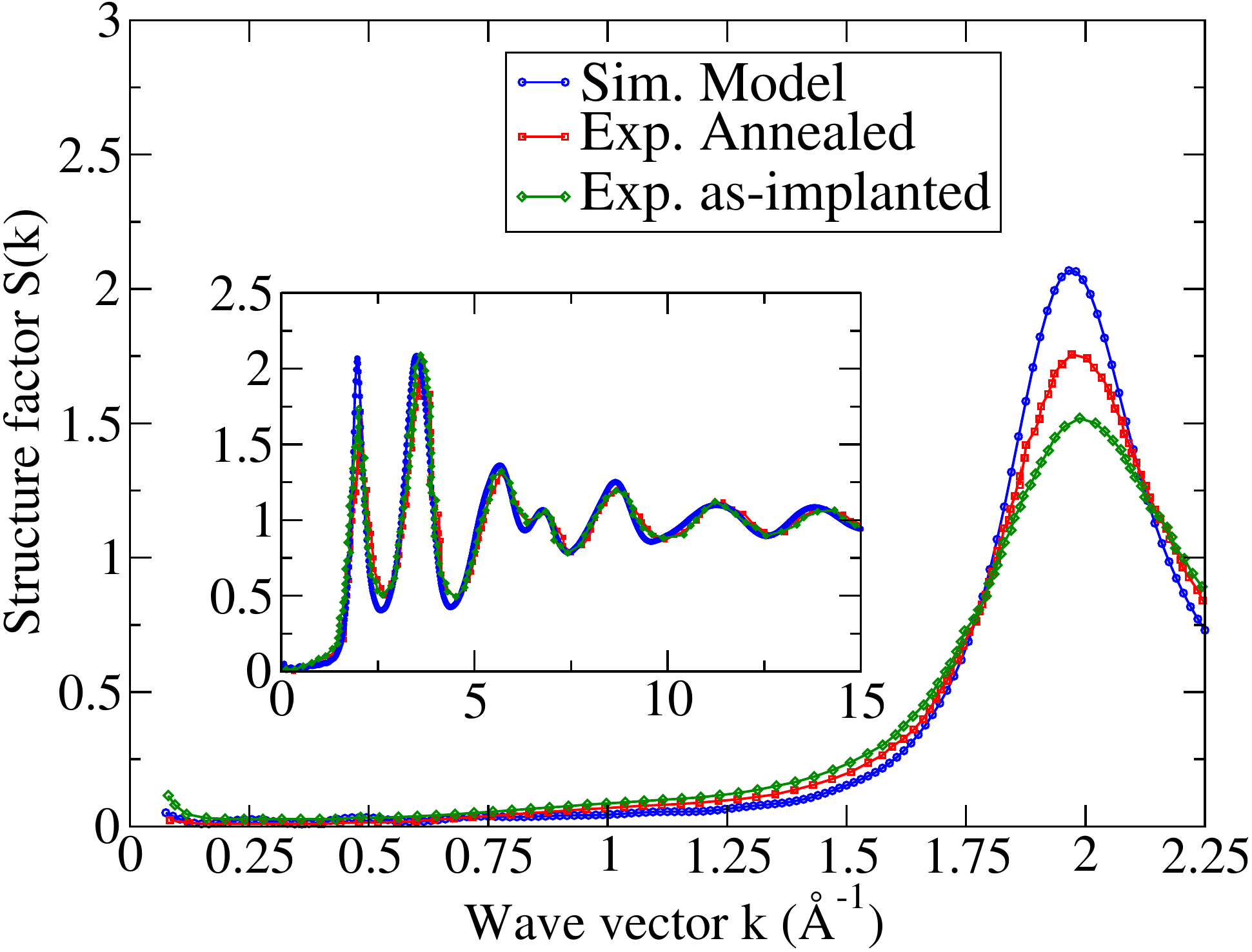}
\caption{
The structure factors of {\asi} obtained from 
experiments and the present simulations. High-resolution 
experimental data at small $k$ for as-implanted and annealed samples 
of {\asi}, from Ref.\,\onlinecite{Xie2013}, are indicated 
in green and red colors, respectively, while the 
simulated data, averaged over M-1 and M-2 configurations, are 
shown in blue.  The full structure factors are presented in 
the inset, with the corresponding experimental data 
(using the same color coding) from Ref.\,\onlinecite{Laaziri1999a}. 
}
\label{FIG3}
\end{figure}
To this end, we shall compute the structure 
factor from model {\asi} networks and compare the same 
with high-resolution experimental structure-factor 
data of {\asi} reported recently in the 
literature.\cite{Xie2013, Laaziri1999a}

\subsection{Structure factor of {\asi} in the small-angle 
scattering region}

In Table \ref{TAB2}, we have listed the characteristic 
structural properties of two models of {\asi}, M-1 and 
M-2, as mentioned earlier in section IIA. Each of 
the models consists of 262,400 atoms in a cubic simulation 
cell of length 176.12 {\AA}, which translates into an 
average mass density of 2.24 g/cm$^{3}$.  The average 
bond angle of 109.23{\deg} between the nearest-neighbor 
atoms is found to be very close to the ideal tetrahedral 
value of 109.47{\deg}, with a root-mean-square 
deviation of $\sim$9.2{\deg}. The average Si-Si bond 
distance is observed to be about 2.39 {\AA}, which is 
slightly higher than the experimental value~\cite{Filipponi1989} 
of 2.36 {\AA} and the theoretical value of 2.38 {\AA} 
reported from {\it ab initio} calculations.~\cite{Car1991}
The number of coordination defects 
is found to be somewhat higher (2.6\%) than the 
values observed in high-quality WWW~\cite{W3-1985}
or ART~\cite{Barkema1996} models obtained from event-based 
simulations but significantly lower than the structural 
models of {\asi} obtained from earlier {\em ab initio} 
and classical molecular-dynamics 
simulations.\cite{Car1991, md_note}
We shall see later in this section that the presence of 
a small percentage of coordination defects, which are 
{\it sparsely} distributed in the models on the atomistic 
length scale of 2--3 {\AA}, do not affect the scattering 
intensity in the long-wavelength limit.

\begin{figure}[t!]
\includegraphics[height=1.6 in, width=1.6 in]{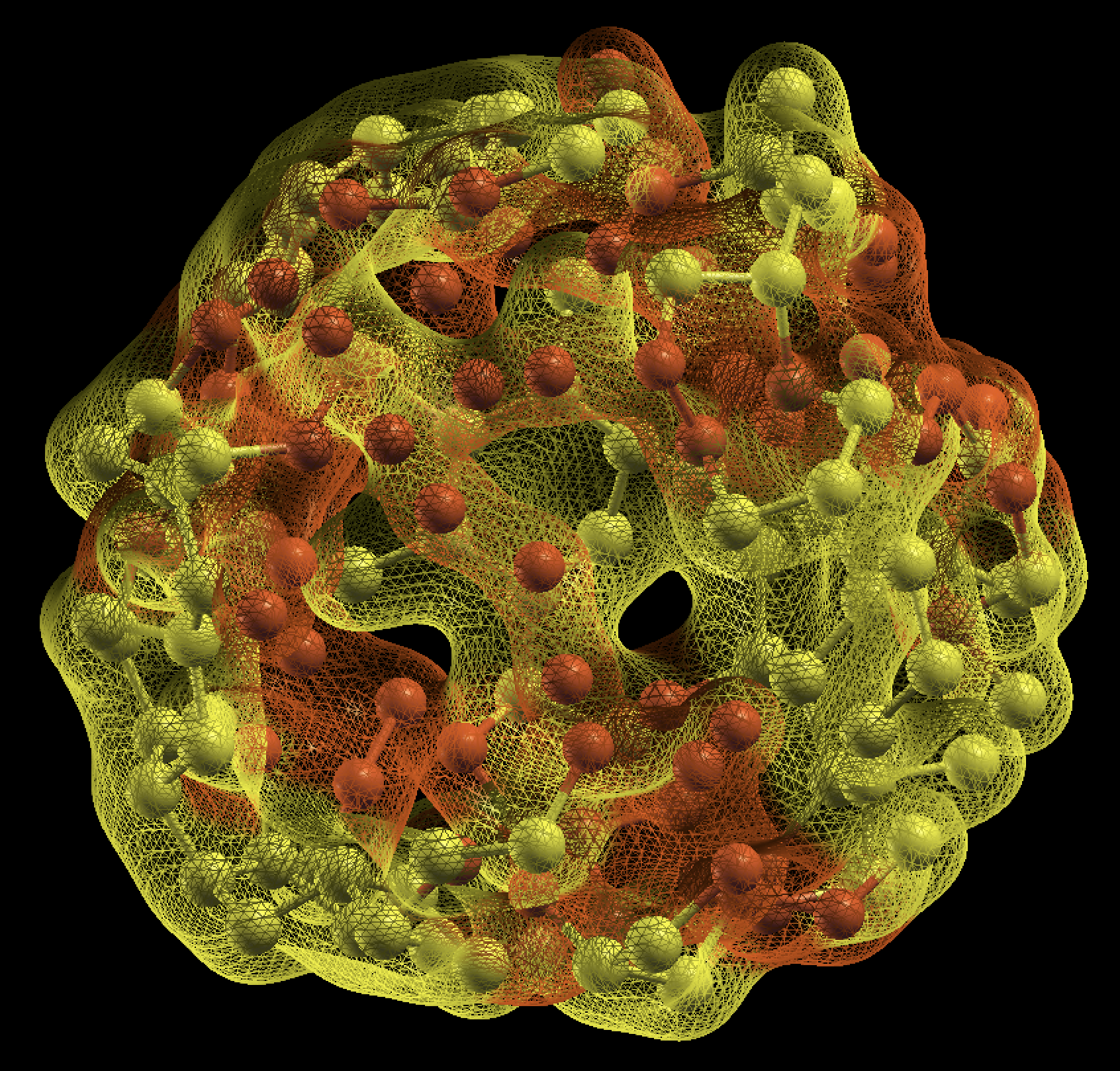}
\includegraphics[height=1.6 in, width=1.6 in]{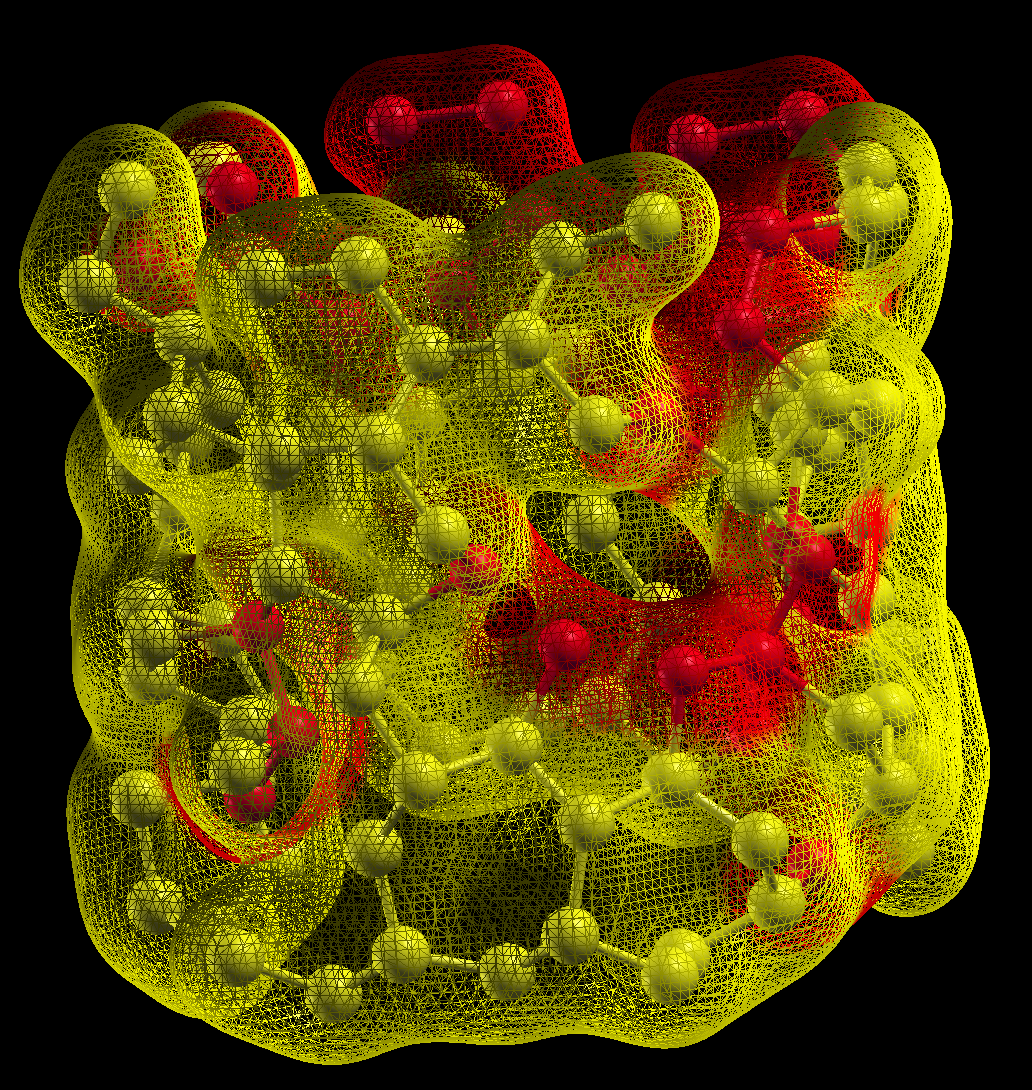}
\caption{ 
Reconstruction of a spherical void surface (in the SP18-R6 model) and a 
cylindrical void surface (in the model CY8-R6 model) during 
total-energy relaxation. For visual clarity, the interface 
atoms within a thin shell of width 2.8 {\AA} and the associated 
reconstructed surfaces are shown in the plot.  The red patches 
on the surfaces indicate the regions which are significantly 
reconstructed via the displacement of Si atoms (red) by 
more than 15\% of the average Si-Si bond length.
}
\label{FIG4}
\end{figure}

\begin{table}[t]
\caption{\label{TAB2} 
Structural properties of {\asi} models used in the present study.
$L$ = Simulation box length (\AA), $\rho$ = mass density (g/cm$^3$), 
$C_4$ = number of four-fold coordinated atoms (\%), $d_{\text{Si}}$ = Average Si-Si 
bond length ({\AA}), $\theta_{\text{avg}}$ = Average bond angle (degree), 
and $\Delta\theta_{\text{RMS}}$ = Root-mean-square deviation (degree). 
}
\begin{ruledtabular}
\begin{tabular}{ccccccccc}
Model & $N$ &  $L$ &  $\rho$ & $C_4$ &  $d_{\text{Si}}$ &  $\theta_{avg}$ &  $\Delta\theta_{\text{rms}}$ \\
\hline
M-1 & 262400 & 176.12 &  2.24 &  97.4 &  2.39 &  109.23{\deg} &  9.26{\deg}      \\ 
M-2 & 262400 & 176.12 &  2.24 &  97.4 &  2.39 &  109.23{\deg} &  9.20{\deg}     \\
\end{tabular}
\end{ruledtabular}
\end{table}        

Having addressed the structural properties of the models, 
we now examine the structure factor, $S(k)$, of {\asi} 
in the small-angle region.  Figure \ref{FIG3} presents 
$S(k)$ obtained by averaging the results from the model 
networks M-1 and M-2. The corresponding experimental data for 
as-implanted and annealed samples of {\asi}, from Ref.\,\onlinecite{Xie2013}, 
are also plotted for comparison. 
Several observations are now in order. First, the simulated 
structure factor agrees well with the experimental 
data obtained from the annealed and as-implanted 
samples for $k$ values up to 15 {\AA}$^{-1}$, as 
shown in the inset of Fig.\,\ref{FIG3}.
Second, an inspection of the simulated and 
experimental data in the vicinity of 1--2 {\AA}$^{-1}$ 
reveals that the former is closer to the annealed data 
than to the as-implanted data. 
This observation is consistent with the expectation 
that {\asi} models from MD simulations should be structurally 
and energetically closer to annealed samples than to 
as-implanted samples. Annealing of as-implanted samples 
at low to moderate temperature (400--500 K) 
reduces the network imperfection locally and thereby 
enhances the local ordering, which reflects in the 
first peak of $S(k)$. 
Third, it is notable that the models have reproduced the 
structure factor in the small-$k$ region, 0.15 $\le$ $k 
\le $ 1 {\AA},$^{-1}$ quite accurately, despite the 
presence of an artificial damping term in Eq.\,(\ref{gr3}) 
that imposes an effective cutoff length of $R_c+5\,\sigma$ 
($\approx$ 35--40 {\AA}) on the radial correlation 
function and the presence of a small number of coordination defects. 

While a direct comparison of the simulated structure 
factor (of {\asi}) with its experimental counterpart establishes the 
efficacy of the numerical approach and the reliability 
of the models used in our study, a more stringent 
test to determine the accuracy of structure-factor data 
in the small-$k$ region follows from the behavior of $S(k)$ in the 
long-wavelength limit. de Graff and Thorpe~\cite{Graff-2010} 
addressed the problem computationally by 
analyzing $S(k)$ as $k \to$ 0, and 
concluded that $S(0)$ was of the order of 0.035 $\pm$ 0.001 by 
studying large {\asi} models containing 10$^5$ atoms. 
Likewise, an analysis of the high-resolution 
experimental structure-factor data of {\asi} 
in the small-angle limit, presented in Fig.\,\ref{FIG3},  
by Xie {\etal}~\cite{Xie2013} 
indicated a value of $S(0) \approx$ 0.0075 $\pm$ 0.0005 from 
experiments. Although a full analysis of the behavior of 
$S(k)$ near $k$ = 0 is outside the scope of the 
present work and will be addressed elsewhere, an 
extrapolation of $S(k)$ at $k$ = 0, by employing a 
second-degree polynomial fit in $k$ in the region 0.15--1.0
{\AA}$^{-1}$, yields a value of 0.0154 $\pm$ 0.0017 in the present study. 
This value 
is comparable to the computed/experimental values 
mentioned earlier and is a reflection of the fact that our models 
produce accurate structure-factor data in the small-angle 
scattering region. The degree of hyperuniformity of a continuous-random-network 
model is often indicated by the value of $S(k)$ at $k$ = 0; a 
low value of $S(0)$ reflects a high degree of 
hyperuniformity.~\cite{Xie2013, Steinhardt2013, Tor1, Graff-2010}

\begin{figure}[t!]
\includegraphics[height=2.75 in, width=2.5 in, angle=270]{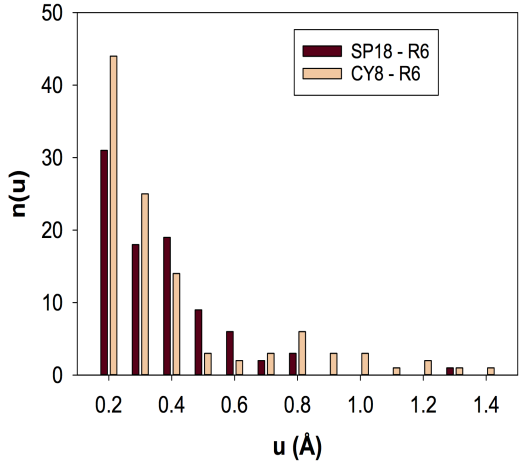}
\caption{
The distribution of atomic displacements ($u$) of the 
interface atoms on a void surface in the models SP18-R6 
and CY8-R6 after total-energy relaxation.  For clarity and 
comparison, only those values of the displacement with 
$u > $ 0.1 {\AA} are shown above. 
}
\label{FIG5}
\end{figure}

\subsection{Reconstruction of void surfaces}
Recent studies on hydrogenated {\asi}, using {\it ab initio} 
density-functional simulations~\cite{BiswasPRA2017, BiswasJAP2014, BiswasIOP2015} 
and experimental data from SAXS,~\cite{Will-JNC-1989} IR,~\cite{Chabal1987, Mahan1989} 
and implanted helium-effusion measurements,~\cite{Beyer2004, Beyer2011} indicate that the shape 
of the voids in {\asih} can be rather complex and that 
it depends on a number of factors, such as the size, number 
density, spatial distribution and the volume fraction of voids, and the method 
of preparation and conditions of the samples/models.  
While the experimental probes can provide considerable structural 
information on voids, it is difficult to infer 
the three-dimensional structure of voids from scattering 
measurements only.  More importantly, experimental data 
from small-angle X-ray and neutron scattering measurements 
include, in general, contributions from an array of 
inhomogeneities with varying shapes and sizes, so it is 
difficult to ascertain the individual role of various 
factors in determining the shape of the measured 
intensity curve in small-angle scattering.  
In contrast, simulation studies are free from such constraints 
and capable of addressing systematically the effect of 
different shapes, sizes, number densities and 
the nature of distributions (e.g., isolated vs.\,interconnected) 
of voids/extended-scale inhomogeneities on scattering intensities. 
Before addressing these important issues, we shall 
first examine the restructuring of a spherical and 
a cylindrical void surface and the resulting changes of its 
shape due to atomic rearrangements on the surface or 
interface region of the voids. 

\begin{figure}[t!]
\includegraphics[height=1.8 in, width=1.8 in]{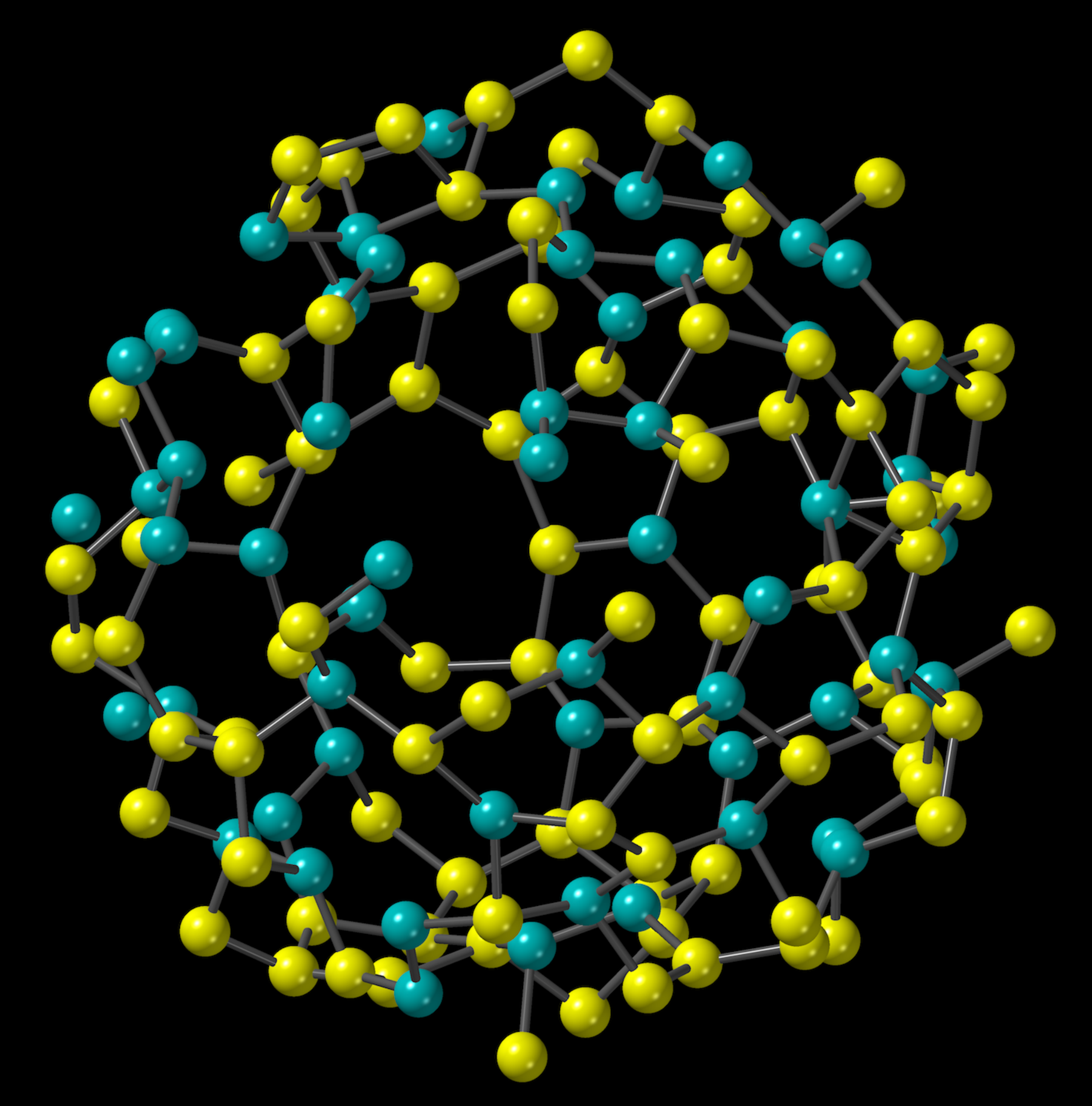}
\caption{Local topological restructuring of a void surface 
(in the SP18-R6 model) via a change of atomic-coordination numbers. 
The silicon atoms, whose coordination number has increased 
[from (2,3) to (3,4)] during the restructuring process are 
indicated in light blue color. 
}
\label{FIG6}
\end{figure}

\begin{figure}[t!]
\includegraphics[width=0.4\textwidth]{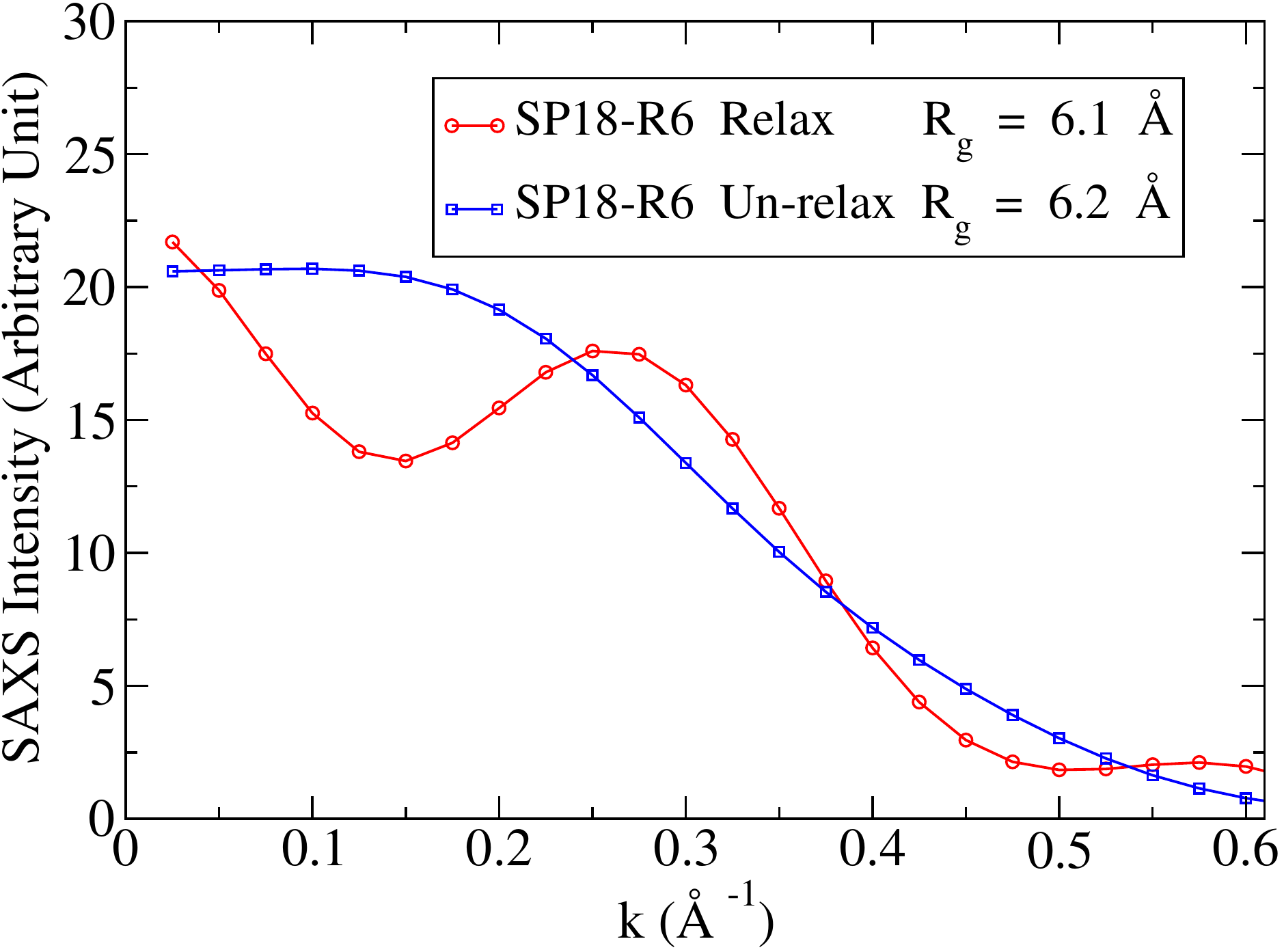}
\caption{
The effect of relaxation on the shape of the scattering 
curve for the model SP18-R6. The variation of the scattering 
intensity results from the three-dimensional restructuring 
of spherical void surfaces, as shown in Fig.\,\ref{FIG4}(left).
}
\label{FIG7}
\end{figure}

Figure \ref{FIG4} shows the reconstructed void surfaces 
of a spherical void of radius 6 {\AA} in the model 
SP18-R6 and a cylindrical void of cross-sectional 
radius 6.1 {\AA} and height 18.3 {\AA} in the model 
CY8-R6. As stated in section IIC, a spherical void is 
defined as an empty cavity of radius $r$ (6 {\AA} for 
SP18-R6) with an interface width $d$ (2.8 {\AA}). 
Atoms within the region between radii $r$ and $r+d$ 
are defined as the surface or interface atoms.  A 
cylindrical cavity or void can be defined in a similar way. 
The radius of gyration of an assembly 
of surface atoms can be readily obtained from the atomic positions 
before and after total-energy relaxation to determine 
the degree of reconstruction and 
the shape of the void. For SP18-R6 and CY8-R6, it has 
been observed that approximately 50\% and 30\% of 
the total surface atoms moved from their original 
position by more than 0.36 {\AA} or 15\% of the 
average Si-Si bond length, respectively, indicating significant 
rearrangements of the surface atoms on the voids. 
A similar observation applies to the rest of the void 
models, where approximately (20--50)\% of the interface 
atoms have been observed to participate in surface 
reconstruction. The interface atoms on a void surface 
in the models SP18-R6 and CY8-R6 are shown in Fig.\,\ref{FIG4} in red colors, 
along with the heavily reconstructed regions of the surface as 
red patches.  The displacement of the interface atoms 
from their original position are 
presented in Fig.\,\ref{FIG5} by showing the distribution 
of the atomic-displacement values.  Such a reconstruction of a void 
surface reduces the strain in the local network 
and increases the local atomic coordination via topological 
rearrangements.  Figure \ref{FIG6} shows several atoms 
(in light blue color) on the surface of a void in model 
SP18-R6, whose coordination number has been found to 
increase from 2--3 to 3--4 upon total-energy relaxation.
The effect of void-surface relaxations on the scattering 
intensity can be readily observed by computing the 
intensity before and after the relaxation. The results for 
the model SP18-R6 are shown in Fig.\,\ref{FIG7}. It is 
apparent that the scattering intensity changes considerably 
upon total-energy relaxation despite the fact that the 
one-dimensional scattering intensity can carry only 
limited information associated with three-dimensional 
structural relaxation of voids.

\subsection{Dependence of SAXS intensity on the size and volume fraction of voids}

Experimental SAXS data on pure and hydrogenated {\asi}
suggest that the scattering intensity in the small-angle 
region is sensitive to the size and the 
total volume fraction of voids present in the 
samples.~\cite{Williamson1995, Mahan1989, Will-JNC-1989, Acco1996}  
Here, we have studied the variation of the scattering intensity 
for different void volume fractions by introducing nanometer-size 
voids of spherical, ellipsoidal, and cylindrical shapes in 
model {\asi} networks. 
Since the scattering intensity from an individual void is 
proportional to the volume of the void, it is necessary 
to choose spherical/ellipsoidal/cylindrical voids of an identical volume 
to ensure that any variation of the intensity can be solely 
attributed to the total volume fraction of the voids. 
\begin{figure}[t!] 
\includegraphics[width=0.4\textwidth]{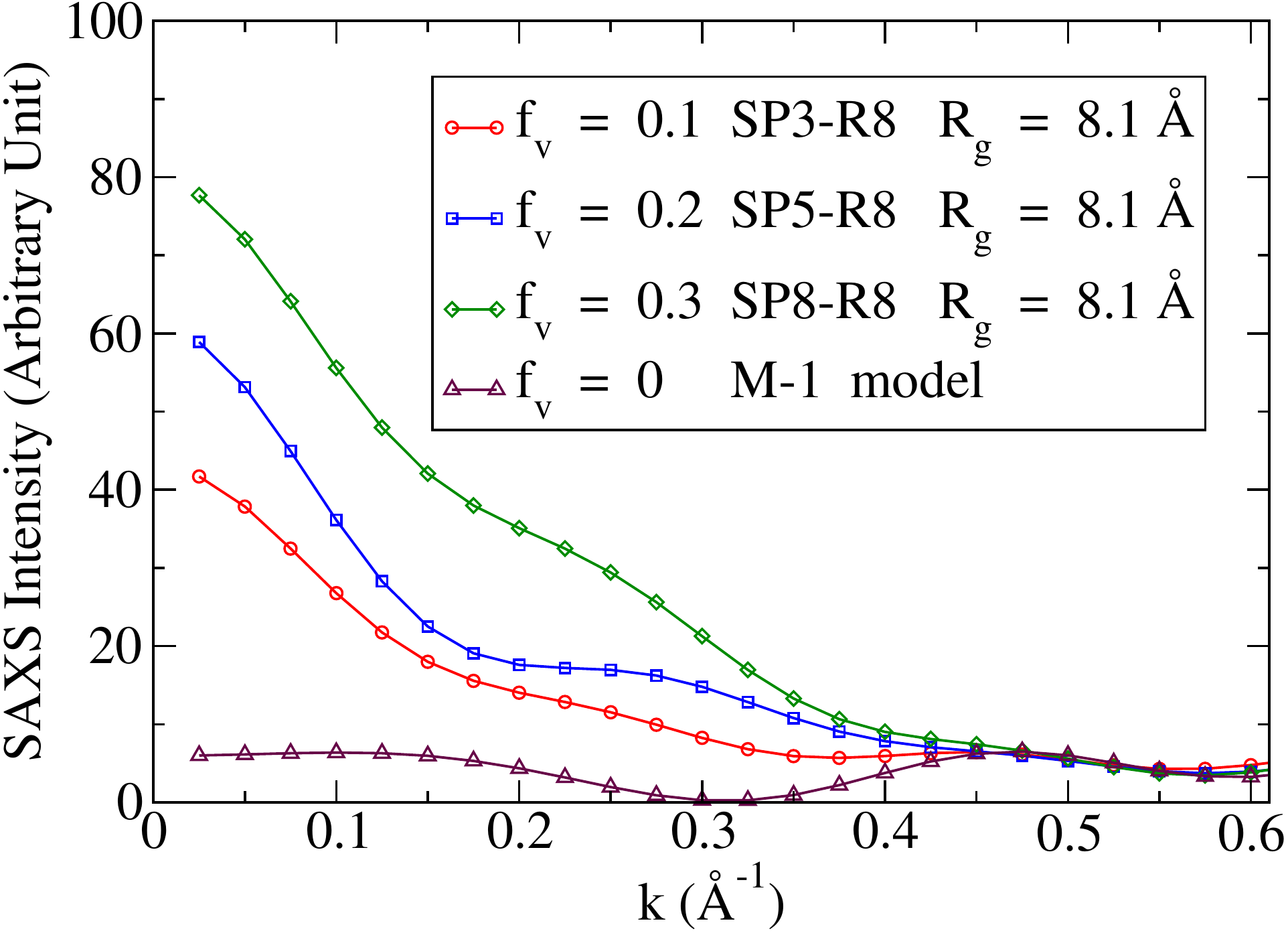}
\caption{
The variation of the scattering intensity (from Eq.\,\ref{EQ6}) 
for four different void-volume fractions.  For comparison, voids of an identical 
shape (i.e., spherical) and size but of different numbers were 
employed in the simulations.  The average radii of gyration of 
the voids are indicated. 
} 
\label{FIG8}
\end{figure}
Following experimental observations,\cite{Williamson1995, NREL411}
we chose void-volume fractions in the range 0.1--0.3\% by 
generating different number of voids of identical volumes 
and shapes. Figure \ref{FIG8} shows the 
intensity variation for four different values of 
the void-volume fraction with an identical 
individual volume of spherical voids.  
For small values of $k$, the scattering intensity strongly 
depends on the volume fraction of the voids and it 
increases steadily with increasing values of the 
void-volume fraction from 0.1\% to 0.3\%.  
Similar observations have been noted for ellipsoidal 
and cylindrical voids but are not shown here.  
%
%A comparison 
%of Fig.\,\ref{FIG6}(a) and Fig.\,\ref{FIG6}(b) suggests that 
%the observation is independent of the shape of the 
%voids as long as the volume of each individual void 
%remains more or less constant.  
%
Likewise, the effect of void sizes on the shape of the 
intensity curve in {\asi} can be addressed in an analogous manner by 
introducing voids of different sizes at a given volume 
fraction of voids.  The results for spherical and cylindrical 
voids for $f_v$ = 0.3\% are presented in Fig.\,\ref{FIG9}. 
An examination of the simulated data presented in Figs.\,\ref{FIG9}(a) 
and  \ref{FIG9}(b) show that there is a noticeable variation in 
the scattering intensity in the small-$k$ region below 0.4 {\AA}$^{-1}$ 
for both spherical and cylindrical voids. 

\begin{figure}[t!] 
\centering
\includegraphics[width=0.4\textwidth]{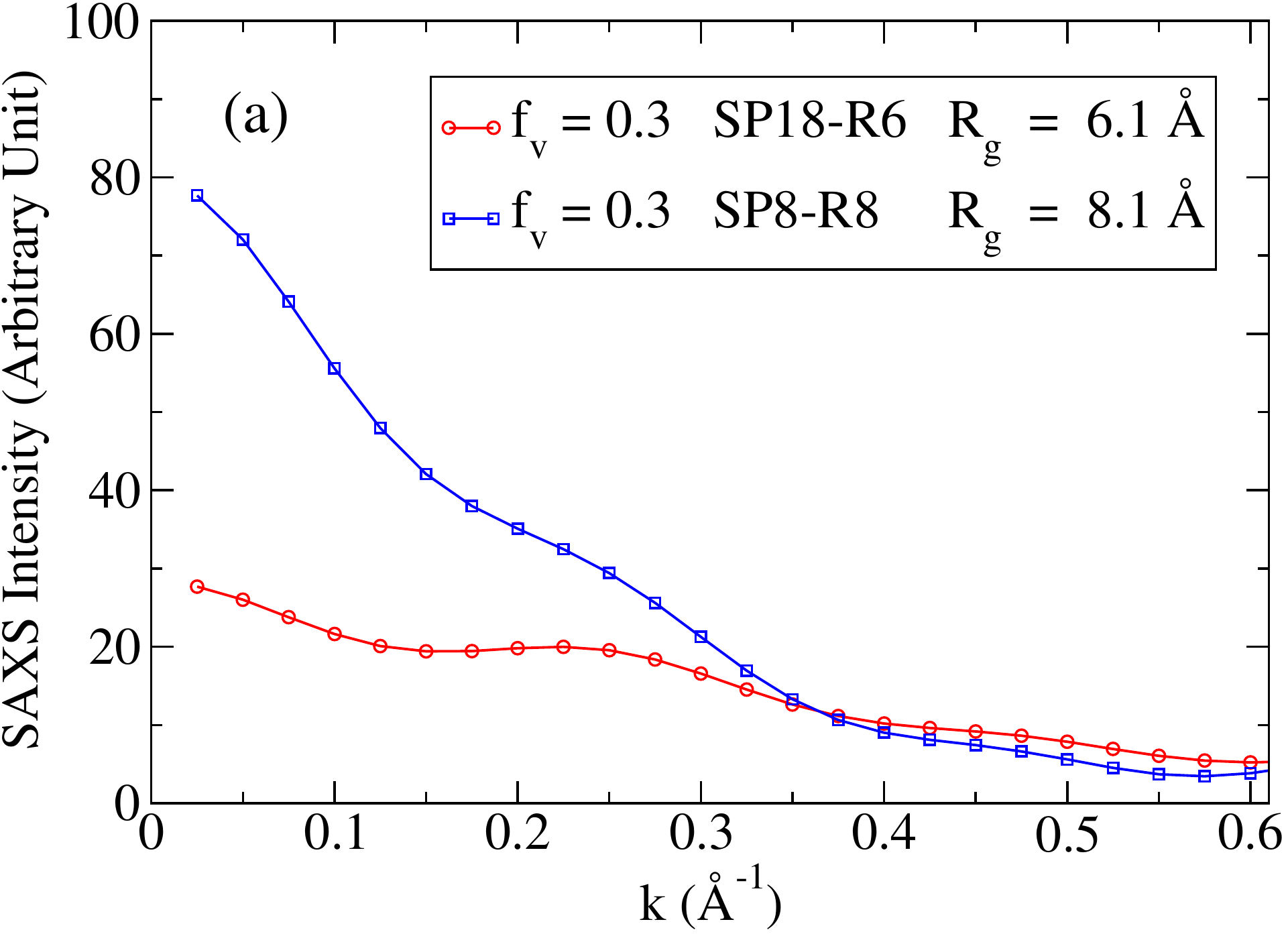}
\includegraphics[width=0.4\textwidth]{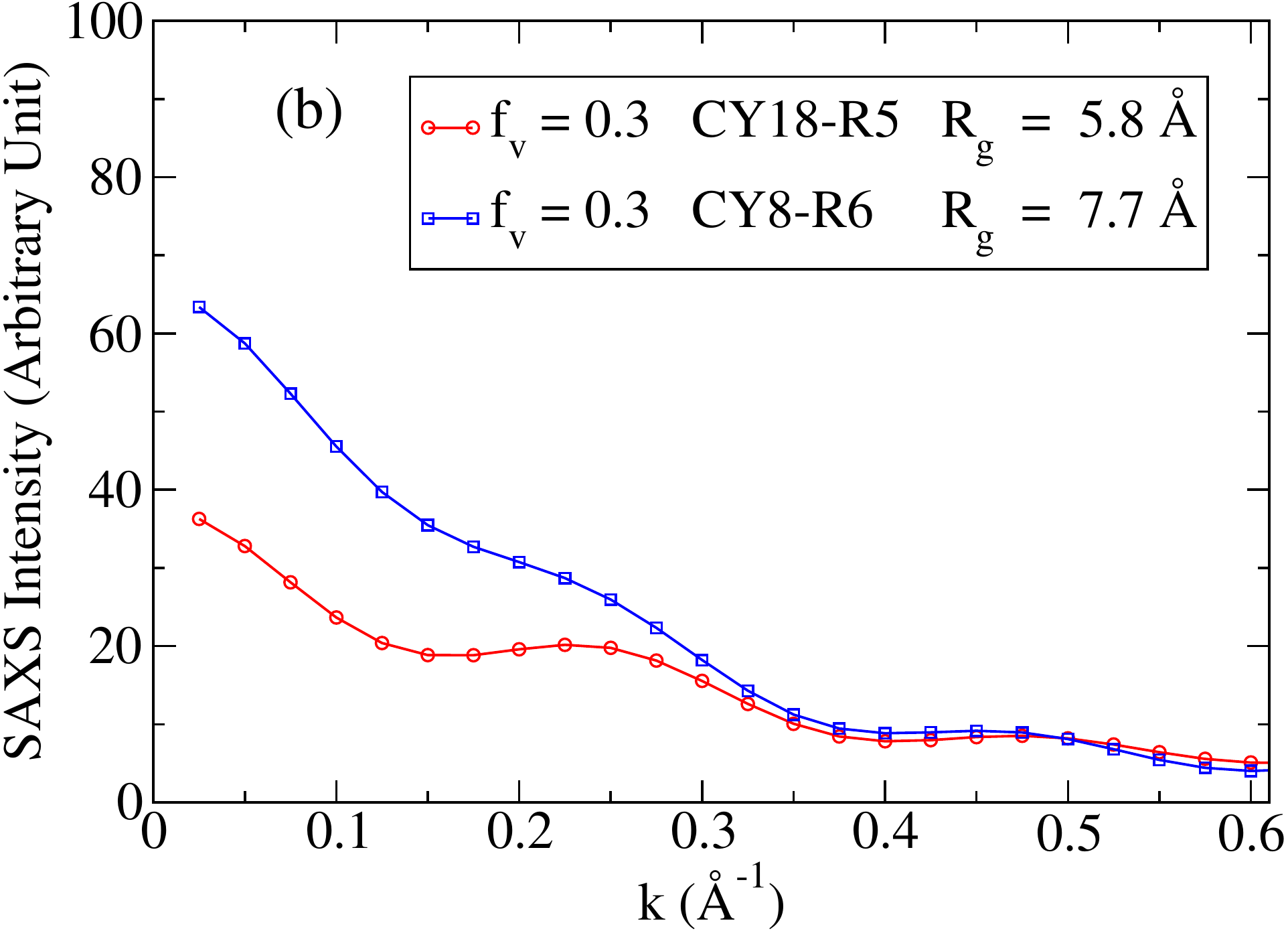}
\caption{
The simulated values of the scattering intensity for two 
different void sizes with 0.3\% volume fraction of the 
voids. 
The results for spherical and cylindrical voids 
are shown in (a) and (b), respectively. The average radii 
of gyration ($R_g$) and the total volume fraction of 
the voids ($f_v$) are indicated in the plots. 
}
\label{FIG9}
\end{figure}

\subsection{Effect of void shapes on SAXS: Kratky plots for {\asi}}

\begin{figure}[t!] 
\centering
\includegraphics[width=0.4\textwidth] {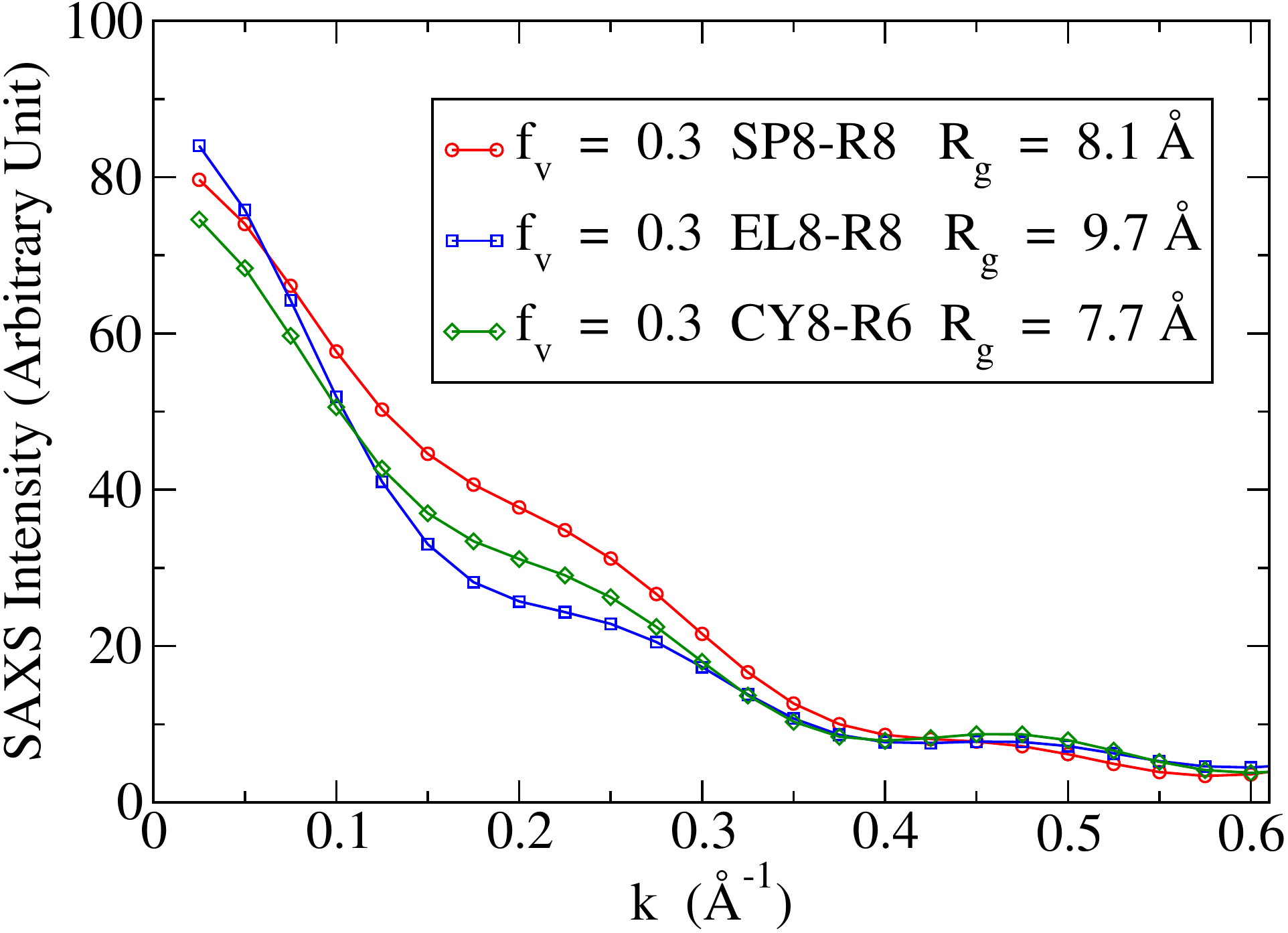}
\caption{
The dependence of the scattering intensity on the shape 
of the voids for a given total volume-fraction of voids 
in {\asi}.  The simulated values of the intensity for spherical (SP), 
ellipsoidal (EL), and cylindrical (CY) voids,  having 
an identical value of the individual void-volume, 
are shown. 
}
\label{FIG10}
\end{figure}

In this section, we have studied the intensity plots for {\asi} 
with spherical, ellipsoidal, and cylindrical voids for an 
identical total volume fraction of the voids to examine the 
effect of the shape and the spatial distribution of the 
voids on the scattering intensity in the small-angle region. 
Since the volume of an individual void can 
affect the scattering intensity considerably, we chose the size 
of the voids in such a way that the individual volumes of the 
voids were identical as far as the total number 
of missing atoms (in a void) is concerned.  
Figure \ref{FIG10} shows the variation of the scattering 
intensities with the wave vector for three models with 
different void shapes, averaged over two independent 
configurations for each model.  Specifically, we have employed 
the models SP8-R8, EL8-R8, and CY8-R6. Each of the models 
contains 8 voids and has a total volume fraction of 
voids of 0.3\%.  Although the average radii of gyration of the voids 
are somewhat different in these models, the individual volume 
of the voids is kept constant to ensure 
that they contribute equally to the total scattering intensity. 
It is evident from Fig.\,\ref{FIG10} that the scattering intensity 
is not particularly sensitive to the shape of the void as 
long as the total volume fraction, individual void volume, and 
the number of voids are identical.  This observation is 
consistent with the earlier experimental studies on {\asih} 
by Mahan {\etal},~\cite{Mahan1989, Mahan1989_solar} 
Leadbetter {\etal},~\cite{Leadbetter1981} and the study 
by Young {\etal},~\cite{Young2007} where a weak dependence 
of the nature of the scattering curve on the shape of the 
voids or inhomogeneities was reported by tilting the incident 
beam with respect to the samples.
In the next paragraph, we will see that a more effective approach 
to determine the effect of void shapes on the scattering 
intensity follows from studying Kratky plots, obtained 
from voids of different shapes.

To examine the relationship between the shape of voids 
and the scattering intensity more closely, we have studied 
the variation of $k^2\,I_c(k)$ with $k$, which 
is often referred to as a Kratky plot in the literature.~\cite{Krat-book} 
Here, following the standard convention in the literature, $I_c(k)$ is 
the background-corrected intensity, which is obtained by 
subtracting the scattering contribution from the amorphous-silicon 
matrix with no voids.  The quantity $k^2I_c(k)$ can be viewed as 
a $k$-space analog of $rG(r)$, which is more sensitive to the intensity variation 
than the conventional intensity $I(k)$, in the same manner as $rG(r)$ is more 
sensitive to structural ordering then the radial pair-correlation function $g(r)$. 
In recent years, Kratky plots 
have  been used extensively in studying the structure of 
biological macromolecules in solution.  It has been 
observed that, for compact and globular (i.e., spherical) 
proteins, the variation of $k^2I_c(k)$ with $k$ is distinctly different 
and stronger than for ones in the partially 
disordered and/or unfolded states.~\cite{Kikhney_2015, Burger_2016} 
Specifically, a globular protein in the folded state exhibits 
an approximate semi-circular variation of $k^2I_c(k)$ with $k$, 
which gradually dissipates or flattens out as the degree of 
structural disorder increases and the protein becomes partially 
disordered by unfolding itself. 
Following this observation, one may expect that the 
shape-dependence of the scattering intensity on a Kratky plot 
would be more pronounced for spherical voids than that for 
long cylindrical or highly elongated ellipsoidal 
voids (see Refs.\,\onlinecite{Mertens2010} and \onlinecite{note5}).

\begin{figure}[t!] 
\centering
\includegraphics[width=0.4\textwidth]{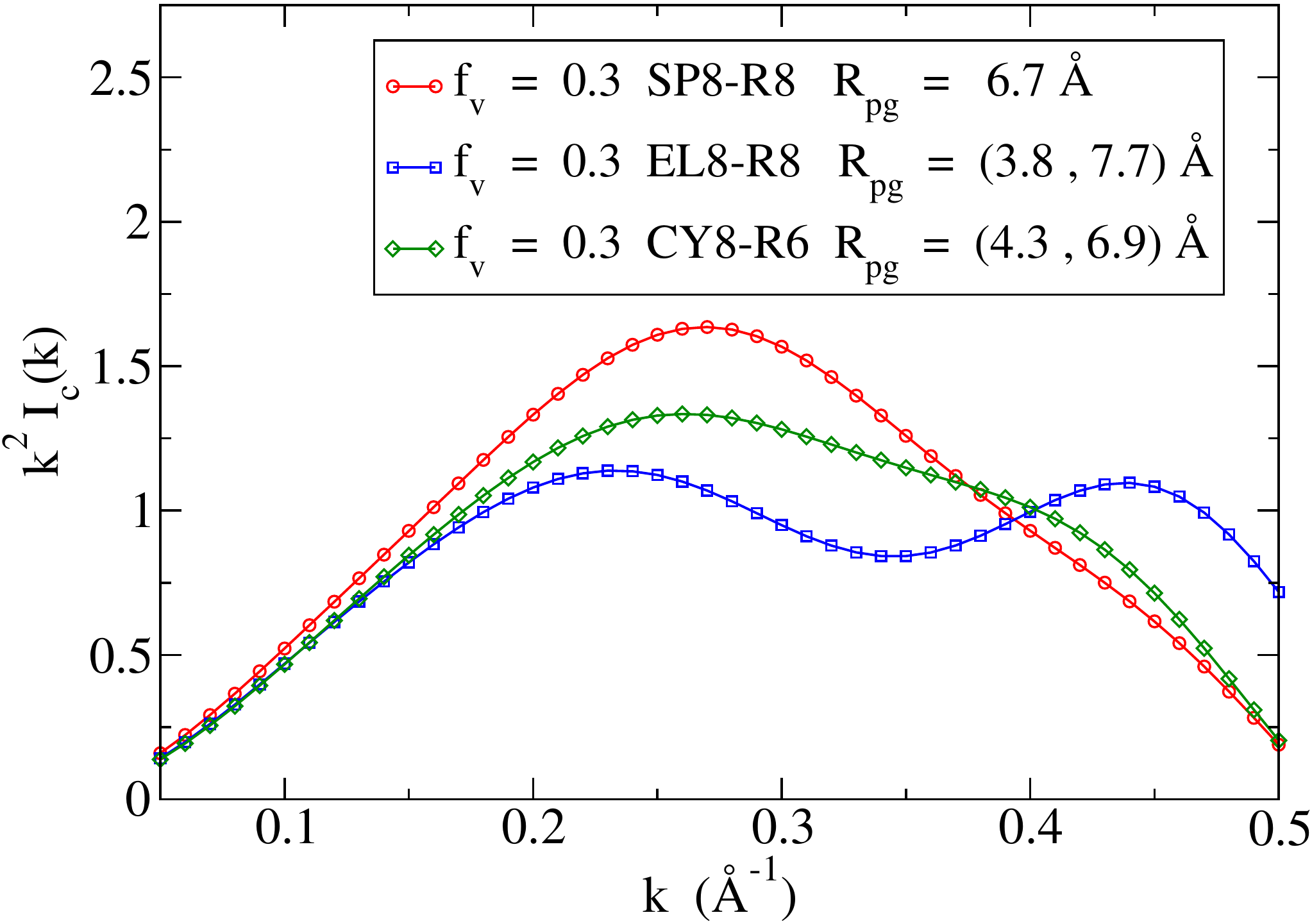}
\caption{ 
Kratky plots showing the variation of the background-correct 
$k^2 I_c(k)$ with $k$ for spherical (SP), ellipsoidal (EL), and cylindrical 
(CY) voids of identical volumes and a total void-volume 
fraction. The pseudo-Guinier radii ($R_{pg}$) 
correspond to the $k$ values obtained from the peak 
position(s) of the intensities for different 
void shapes. 
}
\label{FIG11}
\end{figure}

Figure \ref{FIG11} shows the variation of $k^2\,I_c(k)$ for spherical (SP), 
ellipsoidal (EL), and cylindrical (CY) voids. 
The results can be understood 
qualitatively as follows. Since the largest dimension 
(length) associated with the spherical, ellipsoidal, 
and cylindrical voids are given by $2R$, $4R$, and 
$2.3R$ (see Ref.\,\onlinecite{Krat-note}), respectively, 
where $R$ is the radius of a 
spherical void, it is not unexpected that the intensity 
variation is most pronounced for the spherical voids 
and vice versa for the (elongated) ellipsoidal voids.
Deschamps and De Geuser~\cite{Deschamps_2011} have shown that 
the peak position(s) ($k_{max}$) in a Kratky plot is (are) 
related to the pseudo-Guinier radius, $R_{pg} = \sqrt{3}/k_{max}$, in 
metallic systems, where the particle-size dispersion is 
usually large. The approach has been recently adopted by Claudio 
{\etal}~\cite{Claudio_2014} to estimate the size of silicon 
nanocrystals in bulk nanocrystalline (nc)-doped silicon from 
small-angle neutron-scattering data in order to study 
the effect of nanostructuring on the lattice dynamics 
of nc-doped silicon. Likewise, Diaz {\etal}~\cite{Diaz_2008} 
employed {\it in situ} SAXS for the detection of globular Si 
nanoclusters of size 20-30 {\AA} during silicon film deposition 
by mesoplasma chemical vapor deposition.  
The SAXS intensity profiles obtained by these authors are 
more or less similar to the one obtained by us for the 
spherical voids. The pseudo-Guinier 
radii obtained from the peak positions in the scattering 
intensity for the spherical, ellipsoidal, and cylindrical voids 
are indicated in Fig.\,\ref{FIG11}. 
The pseudo-Guinier radius of 6.7 {\AA},  obtained from 
the Kratky plot in Fig.\,\ref{FIG11}, for the spherical 
voids, matches closely with the initial radius of 8 {\AA} before relaxation. 
For ellipsoidal and cylindrical voids, the presence of two peaks 
is clearly visible in the respective Kratky plots, which 
correspond to linear sizes of (3.8,7.7){\AA} and (4.3,6.9){\AA}, 
respectively.  The presence of multiple peaks in a Kratky plot is indicative 
of a non-spherical shape of scattering objects. 
The lengths associated with these peaks are comparable to the ideal 
values of (4, 8){\AA} (minor and major axes) for ellipsoidal voids 
and (6, 9){\AA} (cross-sectional radius and height) for cylindrical 
voids before relaxation.  We shall see in section 3F that 
the values of the pseudo-Guinier radii are also quite close to the values 
obtained from a conventional Guinier approximation and 
the average radii of gyration computed from the spatial 
distribution of the interface atoms in the vicinity of 
voids in a model. 

\subsection{Effect of spatial distributions of voids on SAXS}

In this section, we address the effect of spatial distributions 
of voids on the shape of the intensity curve in SAXS. Before 
discussing our results, we make the following observation. 
The application of the homogeneous-medium approximation 
in the dilute concentration limit of the inhomogeneities or 
particles, such that the particles are spatially well-separated,  
with a maximum linear size of $l$, suggests that the 
scattering intensity for monodisperse particles solely 
depends upon the volume ($V(l)$), number density ($N(l)$) 
and the shape of the particle for a given density 
difference ($\Delta \rho$) between the particles and 
the average density of the medium.  Following 
Guinier~\cite{Guinier-book} and others,~\cite{Letcher_1966, Vrij_1989, Elliott_book} 
the scattering intensity in this approximation can 
be expressed as, 
\be
I(k) = (\Delta\rho)^2 \, V(l) \, N(l)\int_0^l 4\pi r^2 \gamma_o(r) 
\,\frac{\sin(kr)}{kr} \, dr,  
\label{hm}
\ee
where $\gamma_o(r)$ is a characteristic shape function of the 
particle whose value lies between 0 and 1. The expression 
in Eq.\,(\ref{hm}) suggests that the scattering intensity 
is independent of the atomic-scale structure of the embedding 
medium, provided that the maximum linear size of the particles 
($l$) is significantly larger than the length scale ($R$) 
associated with the atomistic structure of the medium, i.e., 
$l >> R$. Given that $l \approx R \approx$ 10-18 {\AA} in 
the present study, it thus follows that the 
criterion for the homogeneous-medium approximation 
is not satisfied adequately and that a dependence of the 
scattering intensity on the spatial distribution 
of voids may be expected. 

%We now discuss the results of our calculations by noting that, 
%given $l \approx R \approx$ 10-18 {\AA} in the present study, 
%the homogeneous-medium approximation does not adequately 
%satisfy here. 

\begin{figure}[t!] 
\centering
\includegraphics[width=0.4\textwidth]{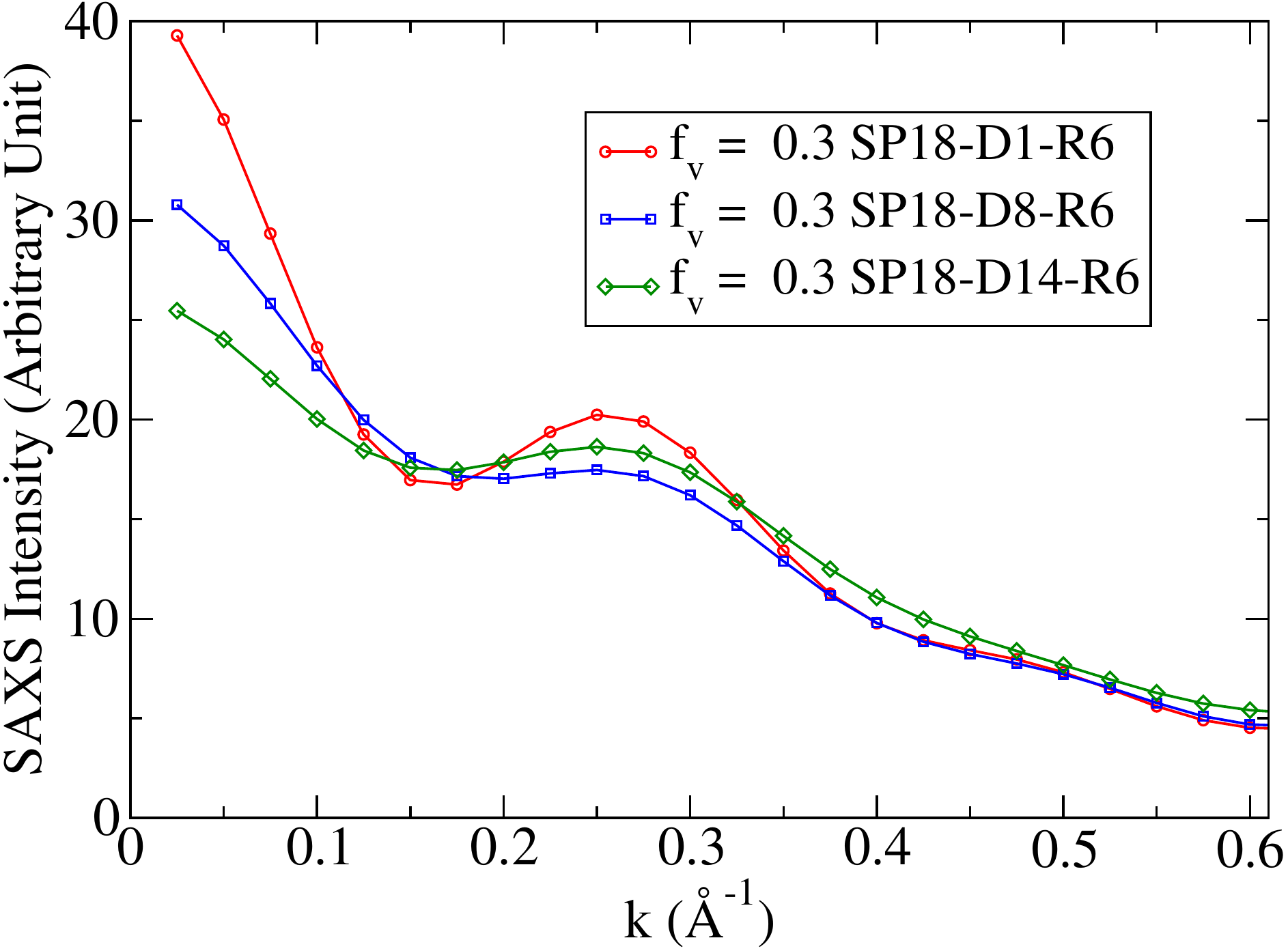}
\caption{
The dependence of the scattering intensity on the spatial 
distribution of the voids in {\asi} for a given volume fraction 
and size of the voids.  The surface-to-surface distance 
($D$) between the voids is indicative of the degree of 
sparseness of the void distribution.  Higher values 
of $D$ correspond to a more scattered or sparse distribution 
of voids. 
}
\label{FIG12}
\end{figure}

The effect of the spatial distribution of the voids on the 
scattering intensity can be studied conveniently by generating 
a number of suitable isolated and clustered distributions 
of voids in real space. 
Since the microstructure of thin-film amorphous silicon is 
characterized by the presence of voids, which cause local 
fluctuations in the (mass) density,  it is important to 
examine to what extent a sparse or interconnected 
distribution of voids can affect the scattering 
intensity in pure and hydrogenated amorphous silicon.  
Using implanted helium-effusion measurements, Beyer 
{\etal}~\cite{Beyer2004, Beyer2011} have shown that the 
presence of He-effusion peaks at low and high temperatures 
are associated with the diffusion of He atoms through an 
interconnected void region and the trapping of He atoms in 
a network of isolated voids, respectively. 
These authors have further 
noted that unhydrogenated samples of {\asi},  prepared by 
vacuum evaporation, can have a high concentration 
of isolated voids. To examine this, we have studied a number 
of models with different spatial distributions of voids. By 
using three different surface-to-surface distances 
($D$ = 1, 8, 14 {\AA}), we have produced three void distributions 
consisting of 18 voids and of radius 6 {\AA}. Each distribution 
corresponds to a volume-fraction density of 0.3\% of voids and is 
reflective of a sparse distribution of voids, as one observes 
in hot-wire or plasma-deposited films of {\asih} at low concentrations
of hydrogen. Figure \ref{FIG12} shows the scattering 
intensity as a function of the wave vector obtained 
for these void distributions.  While it is apparent that the 
intensity is not strongly sensitive to the void distribution, it is 
quite pronounced in the region of $k$ below 0.1 {\AA}$^{-1}$ 
and in the vicinity of 0.26 {\AA}$^{-1}$ for smaller 
values of $D$. A similar observation has been noted for 
the model CY18-R6 but the results are not shown here. 
This dependence can be attributed to the local density fluctutaions 
and the interaction between neighboring voids, which can 
originate from a clustered or interconnected distribution of voids 
produced by a small value of $D$. This is particularly likely in {\asih} at high 
concentrations of hydrogen, where the void distribution has 
been observed to be highly interconnected both from 
experiments~\cite{Beyer2004, Beyer2011} and 
{\it ab initio} simulations.~\cite{BiswasPRA2017, BiswasIOP2015}
However, since the values of the intensity for $k < $ 0.1 
{\AA}$^{-1}$ is sensitive to the numerical noise in $G(r)$ 
and the real-space cutoff $R_c$, it is difficult to 
determine the behavior of the scattering intensity for 
wave vectors below 0.1 {\AA}$^{-1}$. 
Thus, it would not be inappropriate to conclude that 
the scattering intensity is noticeably affected by 
the spatial distribution of voids, especially for a 
sparse distribution, for a given void-volume fraction 
in the small-angle region of $k \ge$ 0.1 {\AA}$^{-1}$.

\subsection{Guinier approximation and the size of the inhomogeneities 
from SAXS}

In writing Eq.\,(\ref{gr1}) from (\ref{xray}) in section IIB, 
we have noted that a peak in $S(k)$,  represented by 
a delta function,\cite{delta_function} at $k = 0$ was excluded 
explicitly to arrive at the expression for the 
static structure factor.  The exclusion of the central peak 
can be readily justified in experiments by recognizing that 
the (central) peak, being dependent on the external shape of the sample, is extremely 
narrow and thus it practically coincides with the incident beam. 
Analogously, one may invoke a similar assumption in the computer 
simulation of SAXS by employing a large but finite-size model of amorphous 
solids so that the computed values of the intensity at small $k$
are minimally affected.
Guinier~\cite{Guinier-book} has shown that, for a homogeneous 
distribution of particles (e.g., voids) in the dilute 
limit, the scattering intensity for small values of $k$ 
can be approximated as,~\cite{Guinier-book}
\be
I(k) = I(0) \exp\left(-\frac{k^2 r_g^2}{3}\right),  
\label{Gu1}
\ee
\noindent provided that the particles are distributed 
randomly with all possible orientations and $kr_g < 1$. 
In Eq.\,(\ref{Gu1}), $r_g$ is the radius of gyration 
of the particles and the inter-particle interaction 
is neglected owing to the dilute nature of their 
distribution. This relationship between the intensity 
and the wave vector in the small-angle limit is widely
known as the Guinier approximation and it is frequently used 
in the experimental determination of the size of 
scattering objects on the nanometer length scale.
The approximation suggests that, as long as the voids are distributed randomly 
(within a large model) in a dilute environment, one 
should be able to estimate the size of the voids from the 
shape of the intensity curve for small values of $k$.  In 
practice, the calculation of the scattering intensity 
from Eq.\,(\ref{Gu1}) is 
constrained by the effective cutoff distance ($R_c$) of 
the reduced pair-correlation function and the size ($l$) 
of the inhomogeneities, which determine the lower and upper 
limits of $k$ in the Guinier approximation, respectively. 
For the present simulations, these values translate to 
an approximate $k$-range from 0.1 {\AA}$^{-1}$  to 
0.5 {\AA}$^{-1}$. 

\begin{figure}[t!] 
\includegraphics[width=0.4\textwidth]{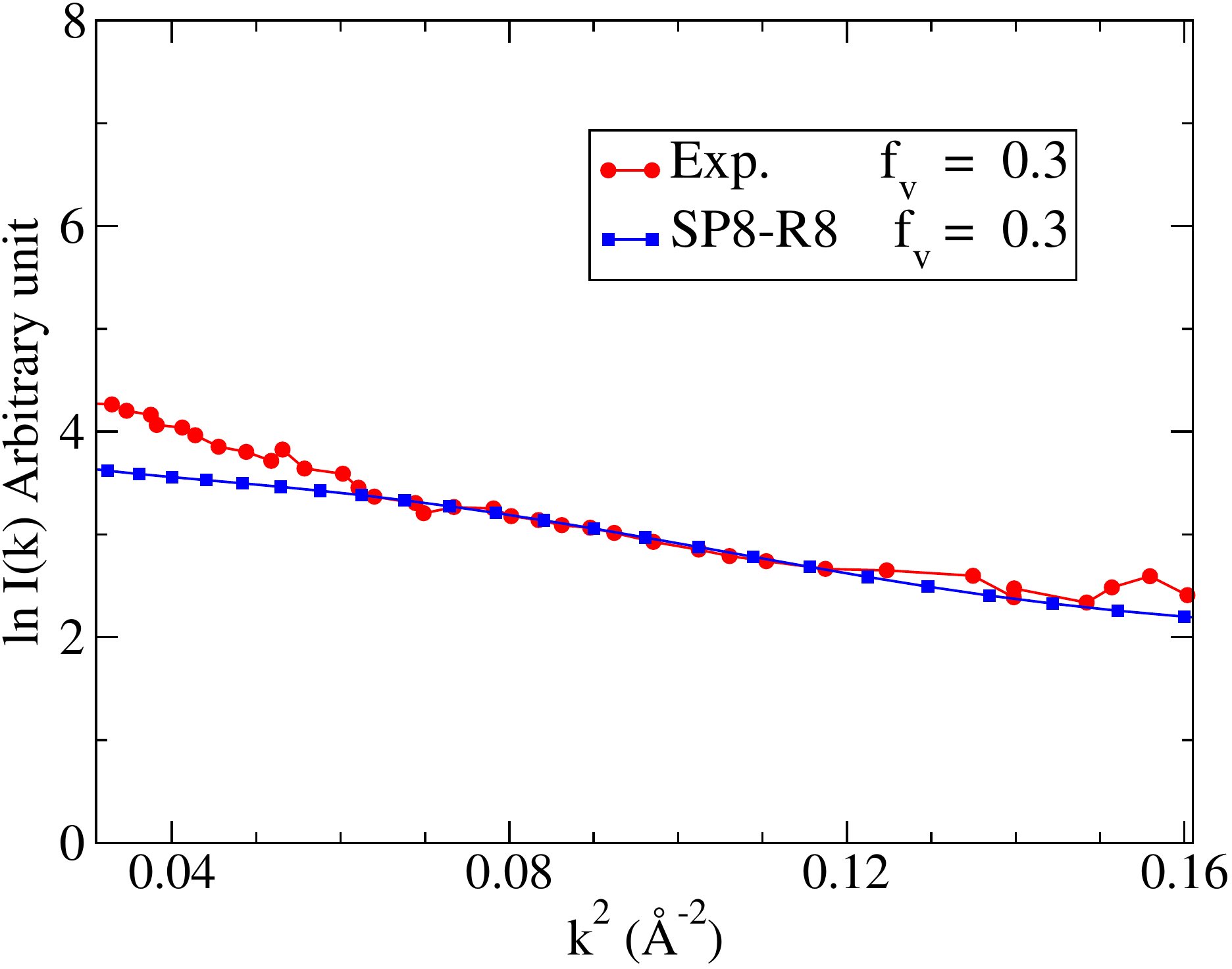}
\caption{
Guinier plots showing a comparison of the experimental 
SAXS data on {\asi}, from Ref.\,\onlinecite{NREL411}, with 
the simulated values for a void-volume fraction of 0.3{\%}. 
}
\label{FIG13}
\end{figure}

\begin{figure}[ht!] 
\includegraphics[width=0.4\textwidth]{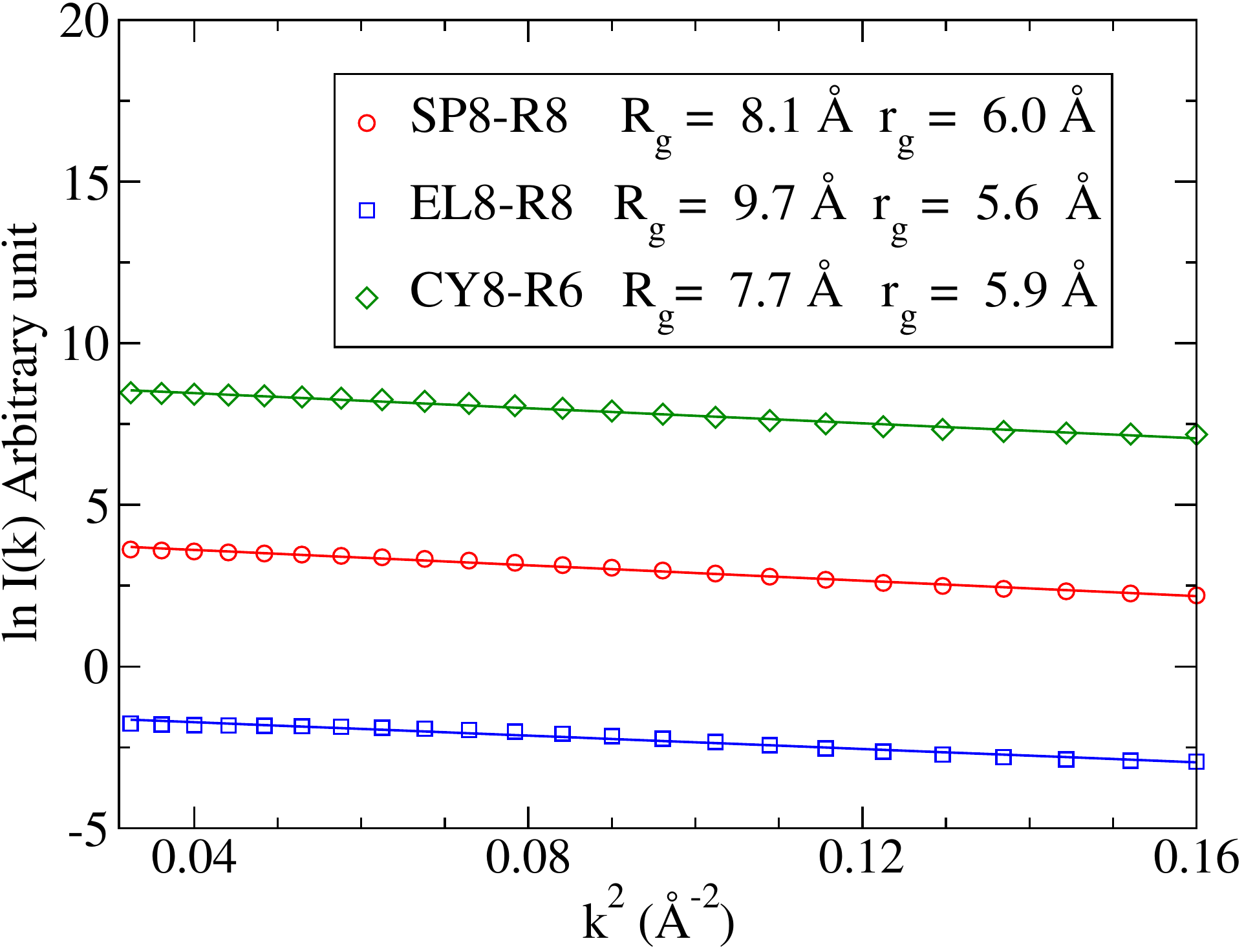}
\caption{
Guinier plots for the simulated values of the 
intensity for spherical (SP), ellipsoidal (EL), and 
cylindrical (CY) voids for a volume fraction of 
0.3\%. $R_g$ and $r_g$ refer to the radius of 
gyration obtained from the distribution of the 
interface atoms and the best-fit Guinier
plots, respectively. For visual clarity, the results 
for the ellipsoidal (blue) and cylindrical (green) voids were given 
a vertical offset of -5 and 5 units, respectively. 
}
\label{FIG14}
\end{figure}

Figure \ref{FIG13} shows a comparison of the experimental 
data from Ref.\,\onlinecite{NREL411} with the results 
obtained from our simulations for a void-volume fraction of 
0.3\% on a Guinier plot, where, following Eq.\,(\ref{Gu1}), 
the scattering intensity is plotted on a natural log scale 
as a function of $k^2$. The simulated values of the intensity 
match closely with the experimental data, except for 
very small values of $k^2$ below 0.05 {\AA}$^{-2}$. The deviation 
for small values of $k$ is not unexpected; it can be 
attributed partly to the difficulty in extracting information 
beyond $R_c$ from the reduced pair-correlation function and 
in part to the intrinsic differences between the simulated models
and experimental samples. Since the latter generally include, 
depending upon the method of preparation and experimental 
conditions, voids of sizes from 5 {\AA} to 15 {\AA}, it 
is difficult to compare simulated data with experimental 
results at a quantitative level for very small values of $k$.
The Guinier approximation in Eq.\,(\ref{Gu1}) suggests that the 
approximate size of the voids/inhomogeneities can be 
obtained from the slope of a $\ln I(k)$ vs. $k^2$ 
plot.  To this end, we have plotted $\ln I(k)$ 
as a function of $k^2$ in Fig.\,\ref{FIG14} for spherical, 
ellipsoidal, and cylindrical voids. Since the values of the 
intensity are close to each other for 
different shapes, the results for the ellipsoidal and 
cylindrical voids are offset by +5 and -5 units, respectively, 
for the clarity of presentation. 
The radii of gyration obtained from the slopes of the fitted 
plots are indicated as $r_g$, whereas $R_g$ reflects the 
average value of the gyrational radius computed from the 
real-space distribution of the interface atoms of 
a void.  Evidently, the latter is larger than 
the actual size of the void. 
For the purpose of comparison, we have subtracted 
1.4 {\AA} -- a length equal to the half of the interface 
width $d$ -- from the value obtained from the Guinier plot 
and have listed the corresponding corrected values for 
each model in the plots and in Table \ref{TAB3}. It may 
be noted that $R_g$ values provide an upper bound of the 
average radius of gyration of the voids, whereas 
$r_g$ values of the same obtained from the Guinier 
plots might have been underestimated in our work 
owing to a possible deviation from the Guinier 
approximation in the scattering region of 
0.1 to 0.6 {\AA}$^{-1}$.  
\section{conclusions}

Small-angle X-ray scattering is a powerful and versatile technique 
for the low-resolution structural characterization of inhomogeneities 
over a length scale of a few nanometers for a variety of ordered 
and disordered materials.  In this work, we have presented a 
computational study of small-angle X-ray scattering in amorphous 
silicon, with particular emphasis on the shape, size, number density, 
total volume fraction, and the spatial distribution of voids in 
amorphous silicon.  Since it is difficult to control these factors 
during experimental sample preparation and hence the analysis 
of the effect of these factors on experimental SAXS data, a direct 
simulation of the scattering intensity is particularly useful in 
studying the variation of the simulated SAXS intensity with respect to these 
factors using atomistic models of amorphous silicon. 
For the accurate simulation of the scattering intensity in the 
small-angle region down to 0.1 {\AA}$^{-1}$, we have produced 
high-quality molecular-dynamical (MD) models
containing 262,400 atoms that correspond to the experimental 
mass density of 2.24 g/cm$^3$ for amorphous silicon. 
The MD models exhibited a narrow bond-angle distribution 
with an average bond angle of 109.23{\deg}$\pm$9.2{\deg} and 97.4\% 
four-fold coordinated atoms. The static structure factors obtained 
from these models agreed quite accurately with high-resolution 
experimental structure-factor data, obtained from transmission 
X-ray scattering measurements. The models exhibited a high-degree of hyperuniformity, 
characterized by the value of $S(k \to 0) \approx 0.0154 \pm 0.0017$, 
which compares well with the value of 0.0075 extracted from the experimental 
structure-factor data. 

An extensive analysis of the simulated SAXS data, obtained by 
varying the size, shape, and the volume fraction of voids introduced 
in the {\asi} models, suggests that the scattering intensity is 
particularly sensitive to the size and the total volume fraction 
of the voids present in the models. The scattering intensity 
increases steadily with an increase of the size of the voids, 
irrespective of the shape and total volume fraction of the 
voids. While the shape dependence is less pronounced in the 
$I(k)$ vs.\:$k$ plots and is consistent with experimental SAXS data, 
an analysis of background-corrected $k^2I_c(k)$ vs.\:$k$ (Kratky) plots for spherical, 
ellipsoidal, and cylindrical voids reveals a clearer picture 
of the overall shape of the voids than the conventional intensity 
versus wave vector plots.  The size of the voids obtained from 
the Guinier approximation and the Kratky plots are more or 
less consistent with each other and comparable with the values
computed from the real-space distribution of the interface 
atoms, provided that the skin depth of the void-surfaces is taken 
into account. 

\section{acknowledgments}
This work was partially supported by the U.S. National Science
Foundation under Grants No. DMR 1507166, No. DMR 1507118, and
No. DMR 1506836. We acknowledge the Texas Advanced Computing 
Center at the University of Texas at Austin for providing 
HPC resources that have contributed to the results reported 
in this work. 

\section*{References}

%\bibliography{ref}
\input{paper4.bbl}

\end {document}

%% file: paper4.bbl

%% file: paper4.bbl
\begin{thebibliography}{77}
\bibitem [{Kra()}]{Krat-note}%
  \BibitemOpen
  \href@noop {} {}\bibinfo {note} {Since the volume of a spherical void and a
  cylindrical void is identical (in the models with an equal number of voids)
  and the height of a cylinder is three times its cross-sectional radius
  $R_{cy}$, this gives $\frac{4\pi}{3}R^3 = 3\pi R_{cy}^3$ or $R_{cy} =
  \sqrt[3]{4/9}R$ and a cylindrical height of $\sqrt[3]{12}R \approx
  2.3R$.}\BibitemShut {Stop}%
\bibitem [{voi()}]{void_atom}%
  \BibitemOpen
  \href@noop {} {}\bibinfo {note} {The term `void atoms' refer to the atoms
  which were removed from the model to form a cavity or void. Likewise, the
  surface atoms indicate those atoms that lie within a spherical shell of
  radius $r$ and $r+d$ for a spherical void. The corresponding surface atoms
  for ellipsoidal and cylindrical voids can be similarly defined.}\BibitemShut
  {Stop}%
\bibitem [{not()}]{note5}%
  \BibitemOpen
  \href@noop {} {}\bibinfo {note} {As far as the shape of scattering objects is
  concerned, the problem of determining the SAXS intensity profile, produced by
  a distribution of compact globular (or spherical) proteins in solution and a
  dilute random distribution of spherical voids in a homogeneous environment,
  can be treated approximately as a primal and its dual problem. In both cases,
  the scattering is due to the difference of mass density of the scattering
  objects from the surrounding medium. Likewise, a partially disordered or an
  unfolded protein -- characterized by a large end-to-end distance or radius of
  gyration -- can be approximated by an elongated scattering object in SAXS.
  Such an approximation is frequenctly employed in coarse-grained
  representation of proteins on a lattice.}\BibitemShut {Stop}%
\bibitem [{del()}]{delta_function}%
  \BibitemOpen
  \href@noop {} {}\bibinfo {note} {For finite-size systems or samples with
  boundary surfaces, it can be shown, following Guinier~\cite{Guinier-book},
  that the central peak at $k$=0 is represented by an integral involving a
  delta function modulated by a shape-dependent term, describing the
  distribution of boundary points. The peak does not contain any structural
  information and it coincides with the incident beam in
  experiments.}\BibitemShut {Stop}%
\end{thebibliography}
